\documentclass[11pt]{article}

\usepackage[final]{acl}

\usepackage{times}
\usepackage{latexsym}

\usepackage[T1]{fontenc}

\usepackage[utf8]{inputenc}

\usepackage{microtype}

\usepackage{inconsolata}

\usepackage{graphicx}

\usepackage{amssymb}
\usepackage{amsmath}
\usepackage{natbib}
\usepackage{graphicx}
\usepackage{url}
\usepackage{multirow}
\usepackage{booktabs}
\usepackage{array}
\usepackage{subcaption}
\usepackage{pifont} 
\usepackage{fix-cm}
\usepackage{authblk}

\title{Beyond A Fixed Seal: Adaptive Stealing Watermark in \\Large Language Models}

\author{
 \textbf{Shuhao Zhang\textsuperscript{1}\thanks{Equal contribution}},
 \textbf{Yuli Chen\textsuperscript{1}\protect\footnotemark[1]},
 \textbf{Jiale Han\textsuperscript{2}\thanks{Corresponding author}},
 \textbf{Bo Cheng\textsuperscript{1}\protect\footnotemark[2]},
 \textbf{Jiabao Ma\textsuperscript{1}}
\\
 \textsuperscript{1}State Key Laboratory of Networking and Switching Technology, 
 \\ Beijing University of Posts and Telecommunications 
 \\ \textsuperscript{2}Hong Kong University of Science and Technology
\\
   \{2020111429, chenyuli\}@bupt.edu.cn, jialehan@ust.hk,\{chengbo, jiabao.m\}@bupt.edu.cn
}

\begin{document}
\maketitle
\begin{abstract}
Watermarking provides a critical safeguard for large language model (LLM) services by facilitating the detection of LLM-generated text. 
Correspondingly, stealing watermark algorithms (SWAs) derive watermark information from watermarked texts generated by victim LLMs to craft highly targeted adversarial attacks, which compromise the reliability of watermarks. 
Existing SWAs rely on fixed strategies, overlooking the non-uniform distribution of stolen watermark information and the dynamic nature of real-world LLM generation processes. 
To address these limitations, we propose Adaptive Stealing (AS), a novel SWA featuring enhanced design flexibility through Position-Based Seal Construction and Adaptive Selection modules. 
AS operates by defining multiple attack perspectives derived from distinct activation states of contextually ordered tokens.
During attack execution, AS dynamically selects the optimal perspective based on watermark compatibility, generation priority, and dynamic generation relevance. 
Our experiments demonstrate that AS significantly increases steal efficiency against target watermarks under identical experimental conditions.
These findings highlight the need for more robust LLM watermarks to withstand potential attacks. 
We release our code to the community for future research\footnote{\url{https://github.com/DrankXs/AdaptiveStealingWatermark}}.

\end{abstract}

\section{Introduction}
\begin{figure}[ht]
  \centering 
  \includegraphics[width=\linewidth]{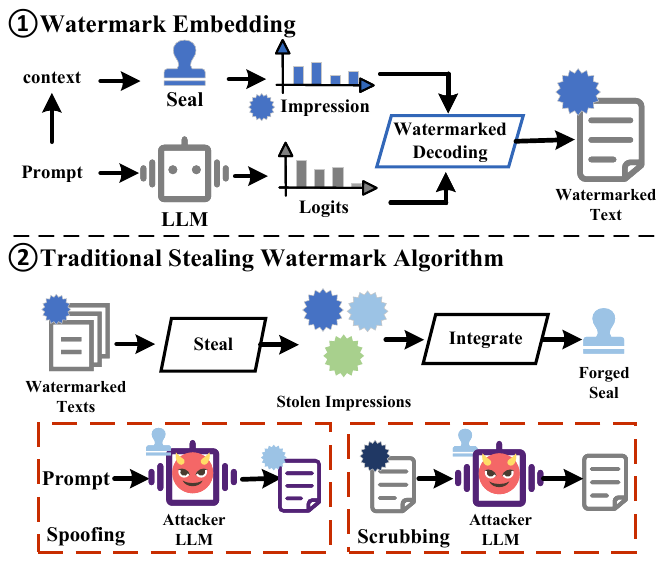}  
  \caption{Part 1 illustrates the watermark generation process in LLMs, while Part 2 depicts the traditional Stealing Watermark Algorithm (SWA).}  
  \label{fig:intro}  
\end{figure}

The proliferation of large language models (LLMs) \cite{MO-GPT4:DBLP:journals/corr/abs-2303-08774,MO-Qwen:qwen2} has introduced significant societal challenges in recent years, including automated phishing \cite{Harm-autoPshing-DBLP:journals/corr/abs-2305-06972}, academic fraud \cite{Harm-academicfraud:labadze2023role}, and misinformation dissemination \cite{Harm-misinformation-DBLP:conf/iclr/ChenS24}. 
While accurately identifying LLM-generated text \cite{Survey-Kath-DBLP:journals/jair/FraserDK25} offers a potential mitigation strategy, the increasing realism of such text renders conventional detection methods ineffective. 
Watermarking \cite{WM2-L-EWD-DBLP:conf/acl/LuLY0K24,WM2-P-UW-DBLP:conf/iclr/HuCWWZH24,WM2-S-SynthID-dathathri2024scalable} emerges as a promising solution: 
As shown in Part 1 of Figure \ref{fig:intro}, the watermarked LLM employs a watermark comprising a \textit{Seal} that generates context-dependent \textit{Impressions}.
These \textit{Impressions} are vocabulary distributions that embed watermark information.
During text generation, watermarks applies these Impressions to guide token selection. 
Later, detection systems verify text origin by measuring the strength of these embedded signals. 

However, watermark reliability faces growing threats from stealing watermark algorithms (SWAs) \cite{AT-SP-stealing-DBLP:conf/icml/0001SV24, AT-SP-Training-DBLP:conf/acl/PanL0LH0KY25}.
By stealing and forging watermarks, attackers can spoof harmful LLM-generated text to falsely appear legitimate, or scrub watermarks from protected texts to evade detection. 
These attacks undermine downstream applications such as content attribution, LLM misuse tracing, and legal accountability \cite{Harm-tracingLLM-Wang2025OnCT,Harm-Authorship:10.1145/3715073.3715076}.
To defend against these threats, proactively studying more advanced SWAs is crucial. Such research is not intended to facilitate attacks, but to uphold the security paradigm of “understanding attacks to build better defenses.”

As illustrated in Part 2 of Figure \ref{fig:intro}, current SWAs forge a fixed seal by statistically extracting impressions from victim watermarked texts. 
\citet{AT-SP-181cww-DBLP:journals/corr/abs-2303-11156} forge a seal targeting KGW's LeftHash-scheme through 2-gram frequency analysis of 181 common words. 
Watermark Stealing (WS) \cite{AT-SP-stealing-DBLP:conf/icml/0001SV24} attacks SelfHash-scheme using three context-processing perspectives, forging and statically weighting multiple seals into one fixed solution. Yet attackers lack knowledge of how victim watermarks process context tokens. 
Since different watermarks assign varying importance to context token positions during impression generation, the inability of fixed seals to dynamically adapt to generation contexts fundamentally limits SWA efficacy against watermarks. 

To overcome this limitation, we propose Adaptive Stealing (AS), a more flexibly designed and effective SWA.
AS introduces two key modules: Position-Based Seal Construction and Adaptive Selection. 
Position-Based Seal Construction systematically generates diverse attack perspectives based on token position activation patterns.
These perspectives facilitate a comprehensive capture of watermark information across various position configurations.
Adaptive Selection dynamically selects the optimal seal from multiple candidate seals based on three criteria: dynamic generation relevance, watermark compatibility, and generation priority. 
The three criteria enable the selected impression to achieve maximal alignment with the watermark patterns in victim texts at each generation step. 
Unlike traditional SWAs that statically integrate multiple perspectives into a fixed seal, AS treats watermark stealing as an adaptive decision process rather than static pattern replication.
Our experiments demonstrate that AS enhances attack effectiveness through the exhaustive utilization of available watermarked texts, highlighting the urgent need for more robust watermarks. 

Our key contributions are as follows: 
\begin{itemize}
    \item We propose Adaptive Stealing (AS), which overcomes the limitations of fixed-strategy attacks through two novel modules: Position-Based Seal Construction, Adaptive Selection. 
     \item We conduct extensive evaluations with real-world attack constraints across diverse watermark configurations. Experimental results demonstrate that AS consistently outperforms the representative baseline under identical conditions and can serve as a powerful tool for evaluating watermarks. 
    \item  Our experiments prove that existing watermarks leak a considerable portion of information. With only 10,000 query samples, AS utilizes the stolen information to almost completely scrubbing three different watermark texts (AUC < 0.55), highlighting the urgent need for more robust watermarks. 

\end{itemize}

\section{Background on LLM Watermarks}
\subsection{LLM Generation} 
Given an autoregressive language model $LM$ and a prompt $[T_1, ...,T_h]$ with $h$ tokens, $LM$ produces a response by generating the next token iteratively.
When generating the token $T_{h+1}$, $LM$ utilizes $[T_1, ...,T_h]$ as input, produces a logit vector $l^{h+1}\in \mathbb{R}^{|V|}$, $V$ is vocabulary of $LM$. 
The logit vector $l^{h+1}$ is then converted into a probability distribution $p^{h+1}$ through softmax.
Then $LM$ samples $T_{h+1}$ from $p^{h+1}$ according to a specific sampling strategy. 
We denote the overall process of decoding the logit vector into a new token as $\mathcal{D}e(\cdot)$

\begin{figure*}[ht]
  \centering 
  \includegraphics[width=0.9\linewidth]{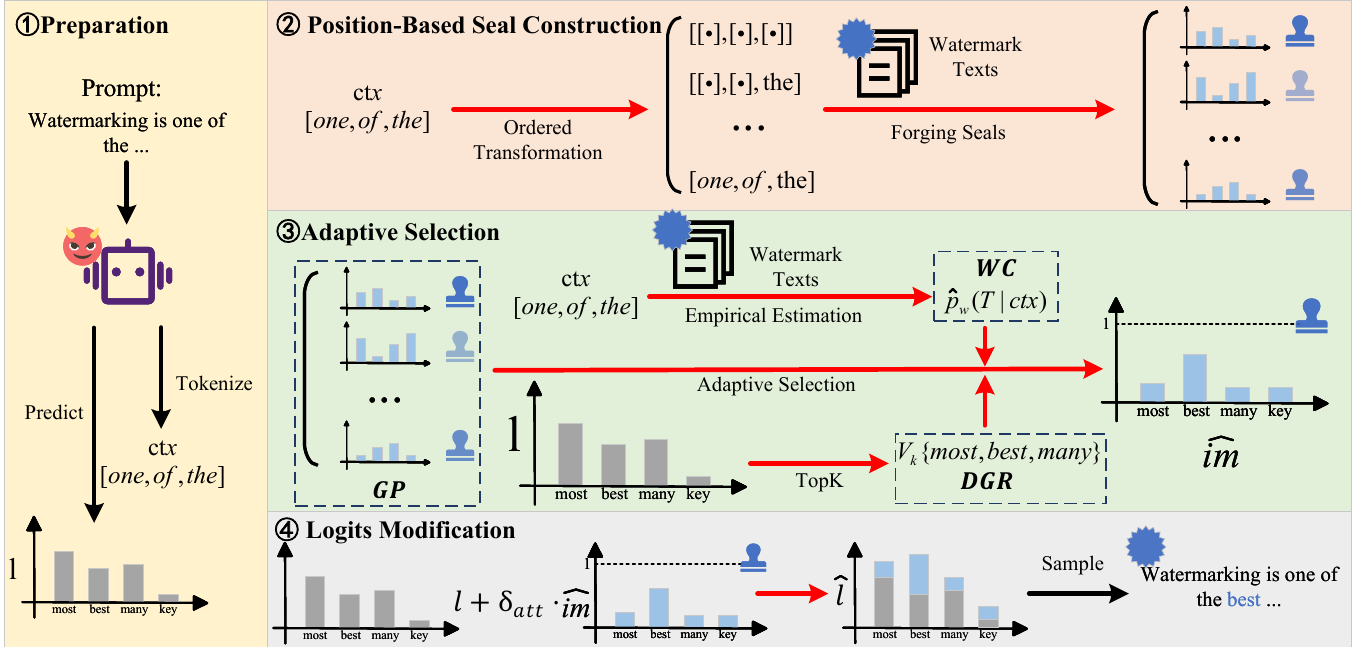}  
  \caption{The overall process of Adaptive Stealing (AS). Red arrows indicate the actions of AS, while black arrows indicate standard generation processes without the intervention of watermarking. }  
  \label{fig:overall-adaptive-stealing} 
\end{figure*}

\subsection{Watermark Embedding}
\label{sec:watermark-embedding}
During the embedding process, a generative watermark utilizes its seal $Seal^\theta(\cdot)$ to generate the impression $im$ dynamically, which is then embedded into the generated text. 
The seal generates the impression by combining $ctx$ and $\mathcal{K}$. 
The context $ctx$ is extracted from the given prompt, defined as the sequence of preceding tokens $[T_{h-|ctx|+1},...,T_{h}]$ of token $T_{h+1}$. 
When generating the impression, $ctx$ introduces dynamics into the process through a predefined Hash-scheme. 
Generally, the Hash-scheme is a part of the seal, and different Hash-schemes are categorized by their approach to processing $ctx$. 
$\mathcal{K}$ is a predefined and private key, which ensures the privacy of watermark. 
Generating impression $im^{h+1}$ at $h+1$ step is formalized as: 
\begin{equation}
\small
    im^{h+1}=Seal^\theta(ctx, \mathcal{K})
\end{equation}
where $\theta$ denotes a set of hyperparameters specific to the watermark. 
$im^{h+1}$ is a vector defined over the $V$ that carries watermark information, serving as a guiding factor to intervene in the generation of token $T_{h+1}$, thereby embedding watermark.

To apply the impression, generative watermarks modify the process $\mathcal{D}e(\cdot)$ by introducing $im^{h+1}$. 
Generating the watermark token $\hat{T}_{h+1}$ by the modified process $\hat{\mathcal{D}e}(\cdot)$ can be formalized as: 
\begin{equation}
\small
    \hat{T}_{h+1}=\hat{\mathcal{D}e}(im^{h+1}, l^{h+1})
    \label{eq:watermark-token-generation}
\end{equation}
Watermarks introduce $im^{h+1}$ through three mainstream approaches.
The first approach is logits vector manipulation, utilizing $im^{h+1}$ to guide the transformation of $l^{h+1}$.
Conversely, the second approach is probability distribution adjustment, which transforms $p^{h+1}$ by $im^{h+1}$. 
The third approach, sampling strategy modification, combines $im^{h+1}$ with the sampling strategy to govern token generation. 
However, regardless of the watermark type, $im^{h+1}$ ensures watermark information, while $l^{h+1}$ ensures the ability of $LM$.

\subsection{Watermark Detection}
\label{sec:watermark-detection}
During the detection phase, the watermark scores a text to determine if it is watermarked. 
First, the watermark detector simulates $Seal^\theta(\cdot)$ to obtain the impression $im$ for each token in the text. 
Then the detector use token-level score function $f_{tls}(\cdot)$ to compute scores for each token based on the corresponding $im$ and $p$ ($p$ is optional for certain watermarks).
Finally, scores are aggregated by a function $Agg(\cdot)$ to obtain Watermark Confidence Score ($WCS$). 
Both $Agg(\cdot)$ and $f_{tls}(\cdot)$ are watermark-specific. 
The detector is formalized as:
\begin{equation}\small
    WCS = Agg(\{f_{tls}(T_i, im^i, p^i)\}_{i=1}^n; \theta)
    \label{eq:WCS}
\end{equation}
where $n$ denotes the length of the text to be detected.
The $WCS$ serves as the final metric to determine whether the text is watermarked. 

\section{Adaptive Stealing}

In this section, we detail traditional SWAs and present our SWA, Adaptive Stealing (AS). 
\textbf{Traditional SWAs} include two steps: forging the seal and applying the impressions. 
In contrast, AS comprises three steps: forging multiple seals, selecting the optimal seal, and applying the impression generated by the selected seal. 
As illustrated in Figure \ref{fig:overall-adaptive-stealing}, we employ \textbf{Position-Based Seal Construction} to forge diverse seals, and select the optimal seal through \textbf{Adaptive Selection}. 
Finally, AS applies the impression via \textbf{Logits Modification}. 

\subsection{Traditional SWAs}
\label{sec:sub-stealing} 

When executing the SWA, an attacker only possesses limited watermarked texts $D_w$ and an assistant language model $LM_{att}$, while having access to unlimited non-watermarked texts $D_n$. 
For a given $ctx$, a higher occurrence count of $T$ in $D_w$ generally implies that the perturbation exerted by the actual watermark impression $im$ is promotive. 
Therefore, the SWA analyzes the occurrence times of ``token $T$ appears after context $ctx$'' in $D_w$ and $D_n$ to steal possible impressions of the victim watermark. 
By integrating these impressions from $D_w$ and $D_n$, the SWA forges a corresponding seal $\hat{Seal}(\cdot)$. 
Specifically, depending on the integration method, different SWAs forge distinct seals. 
However, all forged seals are capable of scoring tokens based on $ctx$, which is formalized as follows:
\begin{equation}\small
    \hat{im}_T=\hat{Seal}(T, ctx, \{D_w, D_n\})
    \label{eq:score-funtion}
\end{equation}
$\hat{im}_T$ is the score for token $T$, and reflects the watermark degree of $T$ given $ctx$. 
By concatenating $\hat{im}_T$ over the entire vocabulary, the impression $\hat{im}$ derived from the forged seal is obtained. 

The approach employed by the SWA to apply the impression parallels the $\hat{\mathcal{D}e}(\cdot)$ process in watermark embedding. 
When providing a prompt to $LM_{att}$, the attacker extracts the $ctx$ from the prompt and then obtains the forged impression $\hat{im}$ via $\hat{Seal}(\cdot)$. 
By configuring distinct directions of applying the impression, the attacker constructs spoofing and scrubbing attacks accordingly.

\subsection{Position-Based Seal Construction} 
The core of Position-Based Seal Construction lies in a critical component of forging the seal, the transformation function $H(\cdot)$. 
$H(\cdot)$ converts $ctx$ into a key $k$, where $k$ retains only a portion of the information from $ctx$. 
When the SWA steals impressions from $D_w$ and $D_n$, $H(\cdot)$ enables different $ctx$ instances to be mapped to the same key $k$. 
This operation mitigates interference from low-frequency $ctx$ instances by aggregating them into the same key $k$ as their high-frequency counterparts. 
A significant portion of sparse $ctx$ inherits the statistical robustness of more frequent ones by $H(\cdot)$. 
Thus $H(\cdot)$ provides them with sufficient statistical significance for the frequency analysis of ``token $T$ appears after context $ctx$'' while stealing impressions. 

For SWAs, $H(\cdot)$ represents a unique attack perspective on the watermark.
It is essentially a hypothesis regarding how $ctx$ is handled within the watermark seal. 
In the watermark, $Seal^\theta(\cdot)$ only activates part of $ctx$ to obtain $im$, paralleling the operation of $H(\cdot)$. 
Therefore, we infer that the transformation produced by the actual seal has a higher correlation with the generated watermark tokens than other potential transformations of $ctx$. 
Since the attacker lacks knowledge of which tokens in $ctx$ the watermark seal activates, we must consider more $H(\cdot)$ to guess how $ctx$ is handled. 

The watermark's $Seal^\theta(\cdot)$ inevitably utilizes the position information of tokens in $ctx$ when handling $ctx$. 
Therefore, we design a set of transformation functions, referred to as ordered transformations, which follow the paradigm $H^o(ctx, n^o)$ to preserve position information.
$k_{n^o}^{H^o}$ denotes the transformation result of $H^o(ctx, n^o)$. 
Here, $n^o$ is an integer satisfying $0 \leq n^o < 2^{|ctx|}$, and represents the activation state of ordered tokens. 
For each $H^o(ctx, n^o)$, $n^o$ is converted into a binary string of length $|ctx|$, serving as the positional labels for the tokens in $ctx$. 
Specifically, a token $T_i\in ctx$ is replaced by a wildcard $[\cdot]$ if its corresponding position label is 0. 
For example, if $ctx = [T_1,T_2,T_3]$ and $n^o=4$, where 4 in binary representation is "100". 
Then, the transformation result is the following:
\begin{equation}\small
    [T_1,[\cdot],[\cdot]]=H^o([T_1,T_2,T_3], 4)
    \label{eq:ordered-transform-example}
\end{equation}
$[T_1,[\cdot],[\cdot]]$ is $k_4^{H^o}$, represents the token at position 1 in $ctx$ is $T_1$, while the tokens at position 2 and 3 are wildcards over the vocabulary.
The transformation result captures the position information of $T_1$, and regards $T_2$ and $T_3$ as inactive.

After setting the length of $ctx$, Position-Based Seal Construction systematically generates $2^{|ctx|}$ possible ordered transformations. 
These ordered transformations maximize the coverage of potential watermark implementations. 
Subsequently, we construct seals based on these ordered transformations. 

Given a $k$ transformed by $H(ctx)$, we obtain conditional distributions $\hat{p}_w(T|k)$ and $\hat{p}_n(T|k)$ through empirical estimation in $D_w$ and $D_n$. 
Subsequently, we follow the work of WS \cite{AT-SP-stealing-DBLP:conf/icml/0001SV24} to define the score function $\mathcal{S}(T,k)$ based on these two conditional distributions:
\begin{equation}\small
    \mathcal{S}(T,k)=\begin{cases}
        \frac{1}{c} \min( \frac{\hat{p}_w(T|k)}{\hat{p}_n(T|k)}, c) & \frac{\hat{p}_w(T|k)}{\hat{p}_n(T|k)} \geq 1 \\
        0 & \text{otherwise.}
    \end{cases}
    \label{eq:as-score-function}
\end{equation} 
$c$ is the hyperparameter to normalize the score to $[0,1]$. 
$\mathcal{S}(\cdot)$ scores the watermark degree of $T$ from the $H(\cdot)$ perspective. 
By concatenating the scores of all tokens in $V$, we obtain the impression for a given $ctx$ from the $H(\cdot)$ perspective. 

Given a fixed $H(\cdot)$, any $ctx$ can be converted into a corresponding impression by Eq. (\ref{eq:as-score-function}), which represents the stolen watermark information. 
All impressions constructed by a fixed $H(\cdot)$ are integrated into a forged seal, which mimics the functionality of $Seal^\theta(\cdot)$ by generating impressions corresponding to different $ctx$. 
Consequently, Position-Based Seal Construction yields $2^{|ctx|}$ distinctly forged seals.

\subsection{Adaptive Selection}
However, when using only Position-Based Seal Construction, directly weighting multiple forged seals as the final seal like WS introduces noise from incorrect seals. 
To address this issue, Adaptive Selection is designed to dynamically select the most adversarial seal at each generation step. 
We posit that most adversarial seal must incorporate three factors: the current generation state, guidance from the available watermarked dataset, and the inherent features of the seal itself. 
Therefore, we propose three design principles for Adaptive Selection: 

\noindent \textbf{Dynamic Generation Relevance (DGR)}: During actual generation, the generation probability of $T$ determines whether we need to consider it when deciding the final token. 
Therefore, when selecting the seal, the impact of low-probability tokens can be disregarded. 
We select the $k$ highest-probability tokens at the current generation step to construct a token set $V_k$:
\begin{equation}\small
    V_{k}=topK(V, p, k)
\end{equation}
$p$ is the distribution generated by $LM_{att}$ at the current step. 
When selecting among multiple forged seals, we only consider tokens in $V_k$. 

\noindent \textbf{Watermark Compatibility (WC)}: The probability distribution $p_w(T|ctx)$ indicates the probability of $T$ generated in the watermark scenario. 
$p_w(T|ctx)$ serves as a direct representation of watermark information and can provide guidance for the selection of seals. 
Specifically, we desire that the impression generated by the finally selected seal maximizes the generation probability of high $p_w(T|ctx)$ tokens. 
Although the real $p_w(T|ctx)$ is unknown, we can replace it with $\hat{p}_w(T|ctx)$, an empirical estimate derived from $D_w$. 

\noindent \textbf{Generation Priority (GP)}: Among all tokens, the normalized score of $\hat{im}_T$ represents the priority of $T$ to be generated when using $\hat{im}$. 
The selection of seals is based on our need to clearly understand which tokens' generation it promotes. 
The naive $\hat{im}_T$ cannot directly represent the promoting effect of $\hat{im}$ on the generation of $T$ because $im$ lacks regularization.
Therefore, for the impression $\hat{im}^{n^o}$ generated by the $n^o$-th forged seal, we transform it into a probability distribution.
We consider the probability of $T$ as the relative significance degree of $T$ in $\hat{im}^{n^o}$. 

After formalizing the three principles, we integrate them to formalize Adaptive Selection as a scoring function $\omega(\cdot)$ for forged seals.
$\omega(\cdot)$ is defined as follows:
\begin{equation}\small
     \omega({n^o},{ctx}) = \underbrace{\sum_{T\in V_{k}}}_{DGR} \underbrace{\vphantom{\sum_{T\in V_{k}}}  \hat{p}_w(T|ctx)}_{WC} \cdot \underbrace{\vphantom{\sum_{T\in V_{k}}} \frac{\hat{im}^{n^o}_T}{\sum_{T'\in V}\hat{im}^{n^o}_{T'}}}_{GP}
    \label{eq:as-weight-s}
\end{equation}
By accounting for the dynamically changing $ctx$ and generation states, Adaptive Selection selects the $n^o$-th seal with the highest $\omega$ score as the final seal.

\subsection{Logits Modification}
Following the work of WS, we apply the impression generated by the final seal $\hat{im}$ by the following formula:
\begin{equation}\small
    \hat{l}=l+\delta_{att} \cdot \hat{im}
    \label{eq:steal-modify}
\end{equation}
$\delta_{att} > 0$ and $\delta_{att}$ is the hyperparameter to control attack strength. 
When executing scrubbing attack, attacker set $\delta_{att} < 0$ in Eq. (\ref{eq:steal-modify}) for watermark removal.

\section{Experiment}
\subsection{Stealing Environment}
The SWA has two roles: the attacker and the victim. 
The victim is a language model $LM_{vic}$ with a watermark $wm$. 
The attacker steals $wm$ from $LM_{vic}$ to construct targeted adversarial attacks.  
However, there are multiple restrictions for the attacker. 

\noindent\textbf{Unknown Parameters}:
The secret key $\mathcal{K}$ and all hyperparameters of the watermark $wm$ are unknown to the attacker. 
Therefore, the watermark information is unattainable for the attacker by a general way. 

\noindent\textbf{Limited Queries}:
The attacker has permission to access $LM_{vic}$ to obtain watermarked responses. 
However, the number of these responses is limited due to financial and time constraints. 

\noindent\textbf{Inaccessible Detection}:
The attacker lacks authorization to access the detection interface of $wm$. 
Free detection access enables attackers to verify victim watermarks, facilitating perfect spoofing/scrubbing attacks.

\begin{table*}[ht]
\centering
\scalebox{0.68}{
\begin{tabular}{@{}cccccccccccccc@{}}
\toprule
                          &                    & \multicolumn{6}{c}{$LM_{vic}$=OPT-2.7b}                                                                 & \multicolumn{6}{c}{$LM_{vic}$=Llama3-8b}                                                                \\ \cmidrule(lr){3-8} \cmidrule(lr){9-14}
                          &                    & \multicolumn{3}{c}{Dolly}                          & \multicolumn{3}{c}{Harm}                           & \multicolumn{3}{c}{Dolly}                          & \multicolumn{3}{c}{Harm}                           \\ \cmidrule(lr){3-5} \cmidrule(lr){6-8} \cmidrule(lr){9-11} \cmidrule(lr){12-14}
Watermark                 & Method             & WCS                               & AUC  & TPR@1\% & WCS                               & AUC  & TPR@1\% & WCS                               & AUC  & TPR@1\% & WCS                               & AUC  & TPR@1\% \\ \midrule
                    -      & Random             & -                       & 0.50 & 0.01    & -                    & 0.50 & 0.01    & -                    & 0.50 & 0.01    & -                      & 0.50 & 0.01    \\ \midrule
\multirow{3}{*}{KGW}                       & Dipper             & \phantom{0}1.225 & 0.75 & 0.05    & \phantom{0}1.207 & 0.75 & 0.06    & \phantom{0}1.585 & 0.80 & 0.17    & \phantom{0}1.650 & 0.80 & 0.27    \\
                          & Dipper+WS          & \phantom{0}0.958 & 0.69 & 0.08    & \phantom{0}1.008 & 0.72 & 0.08    & \phantom{0}0.507 & 0.61 & 0.06    & \phantom{0}0.470 & 0.61 & 0.06    \\
                          & \textbf{Dipper+AS} & -0.002                            & 0.47 & 0.03    & \phantom{0}0.169 & 0.50 & 0.03    & -0.922                            & 0.37 & 0.02    & -0.519                            & 0.40 & 0.03    \\ \midrule
\multirow{3}{*}{SynthID}  & Dipper             & \phantom{0}0.506 & 0.70 & 0.08    & \phantom{0}0.505 & 0.67 & 0.07    & \phantom{0}0.508 & 0.75 & 0.18    & \phantom{0}0.508 & 0.76 & 0.19    \\
                          & Dipper+WS          & \phantom{0}0.502 & 0.57 & 0.04    & \phantom{0}0.502 & 0.55 & 0.02    & \phantom{0}0.503 & 0.62 & 0.05    & \phantom{0}0.503 & 0.62 & 0.05    \\
                          & \textbf{Dipper+AS} & \phantom{0}0.499 & 0.48 & 0.01    & \phantom{0}0.499 & 0.44 & 0.01    & \phantom{0}0.501 & 0.54 & 0.03    & \phantom{0}0.501 & 0.55 & 0.04    \\ \midrule
\multirow{3}{*}{Unbiased} & Dipper             & \phantom{0}0.346 & 0.69 & 0.06    & \phantom{0}0.319 & 0.70 & 0.03    & \phantom{0}0.342 & 0.68 & 0.06    & \phantom{0}0.399 & 0.70 & 0.07    \\
                          & Dipper+WS          & \phantom{0}0.133 & 0.58 & 0.02    & \phantom{0}0.161 & 0.62 & 0.01    & \phantom{0}0.151 & 0.59 & 0.02    & \phantom{0}0.174 & 0.59 & 0.02    \\
                          & \textbf{Dipper+AS} & -0.065                            & 0.47 & 0.01    & -0.050                            & 0.51 & 0.00    & -0.011                            & 0.50 & 0.01    & \phantom{0}0.008 & 0.51 & 0.00    \\ \bottomrule
\end{tabular}
}
\caption{Results of scrubbing attack for different watermarks and different scrubbing method. ``Random'' represents random classification during detection. \textbf{Lower} WCS, AUC, and TPR@1\% values indicate better attack effectiveness.}
\label{tab:scrubbing-table}
\end{table*}

\begin{table*}[ht]
\centering
\scalebox{0.68}{
\begin{tabular}{cccccccccccccc}
\toprule
                          &      & \multicolumn{6}{c}{$LM_{vic}$=OPT-2.7b}               & \multicolumn{6}{c}{$LM_{vic}$=Llama3-8b}              \\ \cmidrule(lr){3-8} \cmidrule(lr){9-14}
                          &                           & \multicolumn{3}{c}{Dolly} & \multicolumn{3}{c}{Harm} & \multicolumn{3}{c}{Dolly} & \multicolumn{3}{c}{Harm} \\ \cmidrule(lr){3-5} \cmidrule(lr){6-8} \cmidrule(lr){9-11} \cmidrule(lr){12-14}
                      Watermark    &     Mode                  & WCS     & AUC   & TPR@1\%  & WCS     & AUC   & TPR@1\% & WCS     & AUC   & TPR@1\%  & WCS     & AUC   & TPR@1\% \\ \midrule
\multirow{3}{*}{KGW}      & w/o Attack                 & 8.880  & 1.00  & 1.00     & 8.933  & 1.00  & 1.00    & 7.991  & 1.00  & 1.00     & 8.072  & 1.00  & 1.00    \\ \cmidrule{2-14}
                          & WS               & 1.228  & 0.74  & 0.06     & 1.404  & 0.77  & 0.11    & 1.333  & 0.74  & 0.15     & 1.431  & 0.75  & 0.19    \\
                          & \textbf{AS(Ours)} & 2.698  & 0.91  & 0.36     & 2.770  & 0.92  & 0.42    & 2.646  & 0.89  & 0.47     & 2.648  & 0.90  & 0.48    \\ \midrule
\multirow{3}{*}{SynthID}  & w/o Attack                 & 0.584  & 1.00  & 1.00     & 0.583  & 1.00  & 1.00    & 0.569  & 1.00  & 1.00     & 0.570  & 1.00  & 1.00    \\ \cmidrule{2-14}
                          & WS               & 0.503  & 0.61  & 0.02     & 0.504  & 0.64  & 0.04    & 0.502  & 0.59  & 0.02     & 0.503  & 0.59  & 0.04    \\
                          & \textbf{AS(Ours)} & 0.505  & 0.68  & 0.06     & 0.507  & 0.73  & 0.13    & 0.504  & 0.63  & 0.04     & 0.504  & 0.66  & 0.06    \\ \midrule
\multirow{3}{*}{Unbiased} & w/o Attack                 & 3.296  & 1.00  & 0.99     & 3.326  & 1.00  & 1.00    & 2.595  & 1.00  & 0.97     & 2.748  & 1.00  & 0.98    \\ \cmidrule{2-14}
                          & WS               & 0.406  & 0.71  & 0.09     & 0.363  & 0.71  & 0.05    & 0.369  & 0.68  & 0.09     & 0.447  & 0.71  & 0.09    \\
                          & \textbf{AS(Ours)} & 0.641  & 0.80  & 0.20     & 0.630  & 0.80  & 0.18    & 0.584  & 0.76  & 0.19     & 0.623  & 0.78  & 0.18    \\ \bottomrule
\end{tabular}
}
\caption{Results of spoofing attacks on different watermarks. w/o Attack represents the detectability performance of victim. \textbf{Higher} WCS, AUC, and TPR@1\% values indicate better attack effectiveness. }
\label{tab:dif-wm}
\end{table*}

\subsection{Settings} 
\textbf{Victim}: 
We select two LLM families for $LM_{vic}$: OPT \cite{MO-opt27-zhang2022opt} widely adopted in watermarking research, and Llama \cite{MO-l3-llama3modelcard}, a prevalent open-source model.
We employ OPT-2.7b and Llama3-8b as two victim LLMs $LM_{vic}$. 
Victim watermarked texts $D_w$ are prepared using the realnews-like subset of C4 \cite{DS-C4:journals/jmlr/RaffelSRLNMZLL20}, which $|D_w|$ is 10,000. 
To demonstrate the superior extraction capability of the AS, the sample size of 10k that we selected is deliberately lower than the sample requirements specified by other SWAs \cite{AT-SP-stealing-DBLP:conf/icml/0001SV24,AT-SP-Training-DBLP:conf/acl/PanL0LH0KY25}. 
The preparation details of $D_w$ are in Appendix \ref{sec:app-expset-prepare-dw}.

We prepare three victim watermarks implemented by Markllm \cite{TOOL-Markllm-DBLP:journals/corr/abs-2405-10051}: KGW\cite{WM2-L-KGW-kirchenbauer2023watermark}, SynthID\cite{WM2-S-SynthID-dathathri2024scalable}, and Unbiased\cite{WM2-P-UW-DBLP:conf/iclr/HuCWWZH24}, which modify logits vector, sampling, and probability distribution, respectively. 
The algorithmic details of victim watermarks are presented in Appendix \ref{sec:app-algs}. 

\noindent \textbf{Attacker}: WS \cite{AT-SP-stealing-DBLP:conf/icml/0001SV24}, the state-of-the-art SWA, is selected as the baseline. More specific attack settings are provided in Appendix \ref{sec:app-expset-attack}. 
Other SWAs like CWS \cite{AT-SP-181cww-DBLP:journals/corr/abs-2303-11156}, WRA \cite{AT-SP-Training-DBLP:conf/acl/PanL0LH0KY25}, and MIP \cite{AT-SP-greenlistPredict-DBLP:conf/acsac/ZhangZZZ0HGP24}, which are less effective in our scenarios, are discussed in Appendix \ref{sec:app-other-swa}.

\noindent \textbf{Evaluation}: 
Following WS, we select prompts from the ``CW'' subset of dataset \textbf{Dolly}\footnote{\url{https://huggingface.co/datasets/databricks/databricks-dolly-15k}} \cite{DS-Dolly:DatabricksBlog2023DollyV2} and construct \textbf{Harm} prompts for harmful content generation by merging parts of HarmfulQ \cite{DS-harmfulq:DBLP:conf/acl/Shaikh0HBY23} and AdvBench \cite{DS-advbench:DBLP:journals/corr/abs-2307-15043} for evaluation. 
For each prompt, we generate 200 tokens by the default attacker model $LM_{att}$ (Qwen2.5-7b) \cite{MO-Qwen:qwen2}. 

Watermarking generally involves a dual-aspect trade-off: detectability and text quality. 
To evaluate detectability, we use the AUC score (Area Under receiver operating characteristic Curve) and TPR@1\% (True Positive Rate at 1\% False Positive Rate). 
We also present the watermark confidence score (WCS) calculated by the corresponding watermark for the detectability evaluation. 
For text quality, we evaluate perplexity (PPL) with the auxiliary model Llama2-13b, which possesses a larger parameter scale than our two $LM_{vic}$.

\begin{table*}[hbpt]
\centering
\scalebox{0.7}{
\begin{tabular}{@{}ccccccccccc@{}}
\toprule
                  & \multicolumn{8}{c}{$n^o$}                                                                         & \multirow{2}{*}{AVE} & \multirow{2}{*}{AS(Ours)} \\ \cmidrule(lr){2-9}
                  & 0     & 1              & 2              & 3              & 4              & 5     & 6     & 7     &                      &                           \\ \midrule
$|ctx|=1$ \& Left & 0.675 & \textbf{0.997} & 0.749          & 0.929          & 0.599          & 0.843 & 0.540 & 0.584 & 0.821                & 0.907                     \\
$|ctx|=2$ \& Left & 0.590 & 0.709          & \textbf{0.994} & 0.945          & 0.649          & 0.635 & 0.855 & 0.692 & 0.834                & 0.864                     \\
$|ctx|=3$ \& Left & 0.560 & 0.559          & 0.680          & 0.694          & \textbf{0.993} & 0.923 & 0.903 & 0.737 & 0.825                & 0.886                     \\
$|ctx|=4$ \& Left & 0.527 & 0.501          & 0.522          & 0.535          & \textbf{0.621} & 0.559 & 0.559 & 0.511 & 0.532                & 0.613                     \\
$|ctx|=3$ \& Min  & 0.692 & 0.940          & 0.915          & 0.932          & 0.888          & 0.860 & 0.845 & 0.753 & 0.850                & \textbf{0.941}            \\
$|ctx|=3$ \& Max  & 0.515 & 0.826          & 0.834          & \textbf{0.875} & 0.810          & 0.753 & 0.754 & 0.685 & 0.730                & 0.846                     \\ \midrule
Unknown           & 0.593 & 0.755          & 0.782          & 0.818          & 0.760          & 0.762 & 0.743 & 0.660 & 0.765                & \textbf{0.843}            \\ \bottomrule
\end{tabular}
}
\caption{Spoofing attack performance (AUC) across different watermark configurations using various seals. $LM_{vic}$=Llama3-8b, and dataset is Dolly. $n^o$ denotes the specific forged seal used for attack, ``AVE'' represents equally weighted ensemble of seals with $n^o$ from 0 to 7. Left, Min and Max respectively indicates LeftHash-scheme, MinHash-scheme and MaxHash-scheme.}
\label{tab:seals-analysis}
\end{table*}

\subsection{Main Results}
As shown in Table \ref{tab:scrubbing-table} and Table \ref{tab:dif-wm}, \textbf{AS consistently exhibits higher attack effectiveness than WS.}
In scrubbing attacks, AS has lower detection metric values than WS. 
The AUC of AS is always below 0.55, while the lowest AUC of WS is 0.55. 
Regarding spoofing attacks, AS also outperforms WS, with consistent AUC improvements of 0.04-0.17. 
Regardless of variations in watermarks, datasets, or victim models, the advantage of AS remains consistent, indicating that AS is a better attack method for evaluating watermark robustness. 

Another noteworthy aspect is that \textbf{AS brings nearly perfect scrubbing attacks.}
In Table \ref{tab:scrubbing-table}, Dipper-processed watermarked texts still retain a significant amount of watermark information, as evidenced by all AUC values for Dipper exceeding 0.65. 
WS enables more targeted processing, removing a greater proportion of the watermark. 
However, AS attains near-optimal AUC values, which are below 0.55, closely approaching the ideal theoretical value of 0.5 associated with random classification. 
The high effectiveness of AS in scrubbing attacks indicates the urgent need for watermarks with higher robustness. 

\begin{figure}[tbp]
  \centering 
  \includegraphics[width=0.95\linewidth]{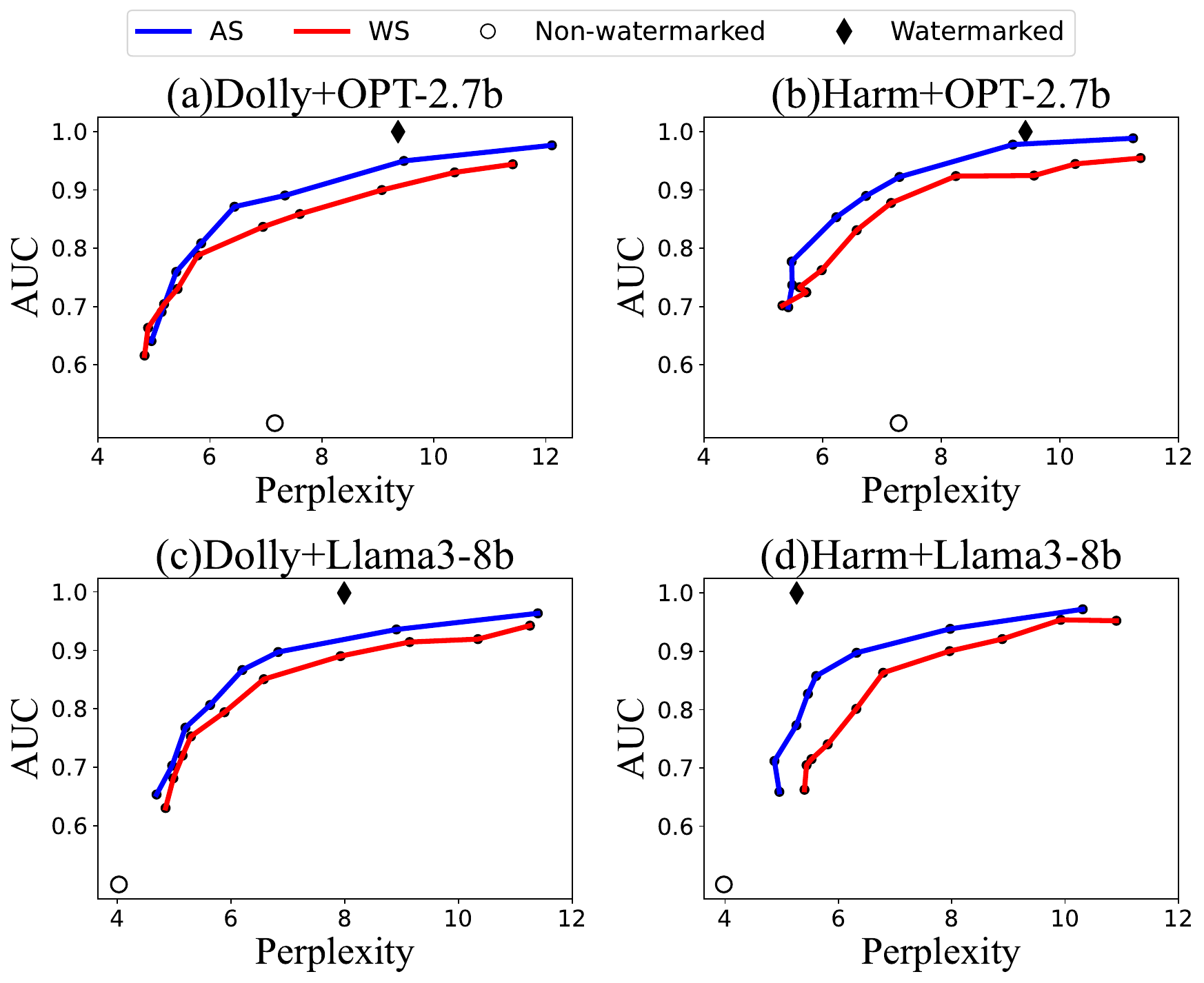}  
  \caption{Trade-off between text quality (PPL) and watermark detectability (WCS) under spoofing attacks on KGW with varying $\delta_{att}$.}
  \label{fig:exp-Textquality}  
\end{figure}
\subsection{Text Quality}

By dynamically adjusting $\delta_{att}$, we analyze the trade-off between detectability and text quality for AS and WS spoofing attacks. 

As shown in Figure \ref{fig:exp-Textquality}, higher $\delta_{att}$ increases AUC but elevates perplexity, indicating higher detectability and reduced text quality.
Compared to WS, AS consistently maintains superior AUC under conditions of high AUC (above 0.75) and comparable perplexity. 
This consistent performance gap at matched perplexity levels indicates that the improvement in attack effectiveness of AS stems from more comprehensive utilization of $D_w$ information rather than text quality degradation.

\subsection{Ablation Analysis}
The two key modules of AS are Position-Based Seal Construction and Adaptive Selection. 
We conduct a separate analysis on the seals generated by Position-Based Seal Construction to demonstrate the roles of both modules in Table \ref{tab:seals-analysis}. 
In this experiment, we configure six distinct seals for KGW, four of which employ LeftHash-scheme with $|ctx|$ ranging from 1 to 4, while the remaining two utilize MinHash-scheme and MaxHash-scheme respectively when $|ctx|=3$. 
``Unknown'' represents the average attack performance across these six victim seals, simulating a realistic scenario where attackers lack knowledge of the watermark's context length and Hash-scheme. 
We also perform an ablation study on the three design principles of Adaptive Selection in Appendix \ref{sec:app-exp-abl-select}. 

Experimental results show that attack performance is optimal when the forged seal configuration matches the victim watermark's parameters.
For example, the seal obtained using $n^o=1$ (corresponding to the leftmost token activation) achieves the highest detection metric when $|ctx|=1$ \& Left, with AUC=0.997 indicating near-complete watermark stealing.
However, when attacking watermarks with different settings, a single seal fails to maintain high aggressiveness. 
Therefore, using Position-Based Seal Construction to obtain seals with diverse attack perspectives is necessary. 

Furthermore, merely acquiring diverse seals does not inherently improve attack performance. 
In Table \ref{tab:seals-analysis}, the equally-weighted AVE approach shows limited improvement over single-seal attacks, failing to leverage the diversity of the forged seals. 
Unlike AVE’s static aggregation, AS adaptively selects the optimal seal for each generation step, leading to superior attack performance. 
Therefore, using Adaptive Selection to flexibly select seals enhances AS's attack effectiveness. 

Through the combination of Position-Based Seal Construction and Adaptive Selection, AS achieves the optimal attack effectiveness in realistic Unknown scenarios (AUC=0.843), consistently outperforming both the single-seal attacks and AVE.

\begin{figure}[tbp]
  \centering 
  \includegraphics[width=0.95\linewidth]{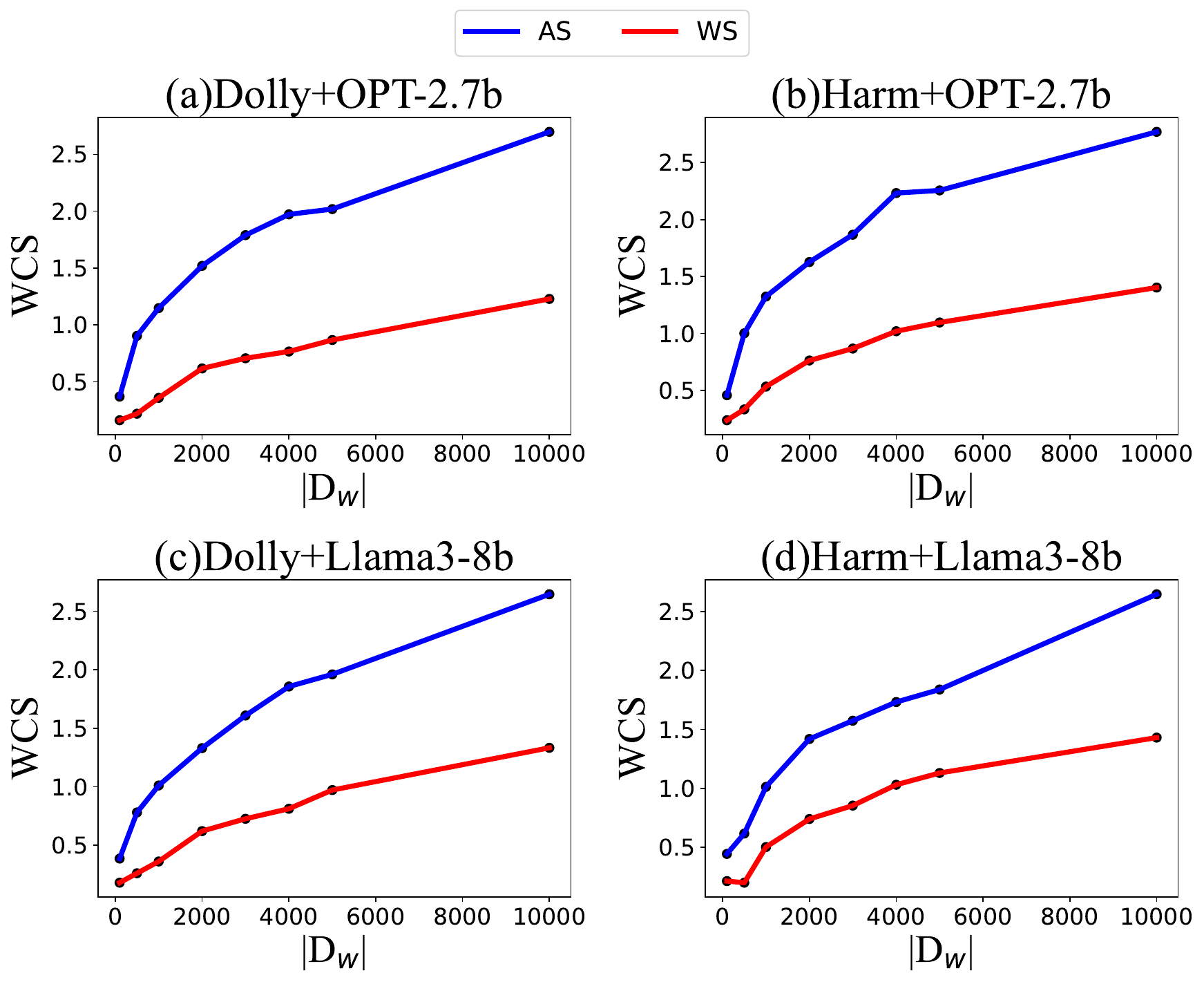}  
  \caption{Results of spoofing attacks on KGW with varying $|D_w|$. The minimum $|D_w|$ is 100.}  
  \label{fig:exp-learnnum}  
\end{figure}

\subsection{\texorpdfstring{$|D_w|$}{|D_w|} Analysis}

A key to SWA lies in the size of victim watermarked texts $D_w$. 
We prepare different $|D_w|$ for the evaluation. 
The results are shown in Figure \ref{fig:exp-learnnum}. 

Both AS and WS demonstrate marked improvements in spoofing performance as $|D_w|$ grows. 
As illustrated in Figure \ref{fig:exp-learnnum}, AS consistently achieves higher WCS values than WS across different $|D_w|$. 
Notably, AS only requires 2,000 victim watermarked texts to achieve attack effectiveness equivalent to that of WS using 10,000 samples. 

Both $LM_{att}$ and $|ctx|$ re equally key factors influencing SWA.
We further analysis $LM_{att}$ and $|ctx|$ in Appendices \ref{sec:app-exp-LM_att} and \ref{sec:app-exp-ctx}, respectively.

\section{Related Work} 
\subsection{LLM Watermarks}
Embedding watermarks into LLM emerges as a promising approach for identifying LLM-generated text \cite{Survey-Kath-DBLP:journals/jair/FraserDK25, Survey-liu-DBLP:journals/csur/LiuPLLHZWKXY25,WM1-EaaS-DBLP:conf/acl/PengYWWZLJXSX23,WM1-RL-DBLP:journals/corr/abs-2403-10553,WM3-Lexical22,WM3-Cater-DBLP:conf/nips/HeXZLWLJ22,WM3-Lexical-DBLP:conf/aaai/HeXLWW22}. 
Current watermarks primarily modify token generation process through three distinct mechanisms: logits vector manipulation \cite{WM2-L-KGW-kirchenbauer2023watermark, WM2-L-SW-DBLP:conf/aaai/FuXD24, WM2-L-SIR-DBLP:conf/iclr/LiuPHM024, WM2-L-Unigram-DBLP:conf/iclr/ZhaoA0W24, WM2-L-XSIR-DBLP:conf/acl/0002ZHL0T0W24,WM2-L-EWD-DBLP:conf/acl/LuLY0K24,WM2-L-E2E-DBLP:conf/icml/WongZ0S25}, probability distribution adjustment \cite{WM2-P-UW-DBLP:conf/iclr/HuCWWZH24, WM2-P-Ruibo-DBLP:journals/corr/abs-2502-11268,WM2-P-BiMark-DBLP:conf/icml/Feng0ZZP25}, and sampling strategy modification \cite{WM2-S-SynthID-dathathri2024scalable, WM2-S-EXP-DBLP:journals/tmlr/KuditipudiTHL24,WM2-S-Christ-DBLP:conf/colt/ChristGZ24}.  
These generative watermarks offer significant practical advantages for LLM services as they eliminate the need for model retraining and incur minimal resource consumption during deployment. 
This efficiency makes them particularly suitable for real-world applications where computational resources are constrained.
\subsection{Watermark Attacks}
Watermark attack methods has two classes: one is scrubbing attacks that remove watermarks from text, and the other is spoofing attacks that forge watermarked text under unauthorized conditions. 
Scrubbing attacks can be implemented using pure text modification techniques\cite{AT-SC-sand-DBLP:journals/iacr/ZhangEFVAB23,AT-SC-DIPPER-DBLP:conf/nips/KrishnaSKWI23,AT-SC-Smoothing:DBLP:conf/emnlp/ChangHS25}. 
Among these, when scrubbing attacks use LLM rewriting, they can use watermark information to achieve more targeted scrubbing \cite{AT-SP-stealing-DBLP:conf/icml/0001SV24}.
Spoofing attacks, in contrast, must utilize watermark information for assistance \cite{AT-SP-181cww-DBLP:journals/corr/abs-2303-11156,AT-SP-stealing-DBLP:conf/icml/0001SV24}. 
The watermark information required by attackers can only be obtained by reverse engineering the Stealing Watermark Algorithm. 
\subsection{Stealing Watermark Algorithm} 
Stealing Watermark Algorithms (SWAs) facilitate two primary adversarial spoofing and scrubbing attacks against watermarked LLMs. 
Extensive research has identified these attacks significant threats to the watermark reliability \cite{WM2-L-SIR-DBLP:conf/iclr/LiuPHM024,Survey-liu-DBLP:journals/csur/LiuPLLHZWKXY25,AT-SP-learnable-DBLP:conf/iclr/GuLLH24, AT-SP-Training-DBLP:conf/acl/PanL0LH0KY25}. 
Existing SWAs, such as those proposed by \citet{AT-SP-181cww-DBLP:journals/corr/abs-2303-11156}, \citet{AT-SP-stealing-DBLP:conf/icml/0001SV24}, \citet{AT-SP-greenlistPredict-DBLP:conf/acsac/ZhangZZZ0HGP24} focus predominantly on compromising KGW variants by forging a fixed seal. 
However, a single static seal is incapable of adapting to diverse watermark implementations or generation contexts. 
We argue that effective SWAs must evolve beyond rigid designs toward adaptive frameworks that dynamically select optimal attack strategies based on contextual requirements.

\section{Conclusion}
We propose Adaptive Stealing (AS), a novel stealing watermark algorithm. 
The key design of AS is Position-Based Seal Construction and Adaptive Selection. 
AS constructs multiple seals based on Position-Based Seal Construction, and Adaptive Selection dynamically selects the most suitable seal to facilitate adversarial attacks. 
AS enhances attack effectiveness by comprehensive stealing of watermark information and precise filtering. 
During the experimental phase, we configure multiple attack environments to evaluate the performance of AS. 
Our experimental results demonstrate that AS consistently exhibits superior attack effectiveness across various attack scenarios. 
These findings demonstrate AS's practical threat value, indicating that current watermark designs still require more robust research to counter potential attacks.

\section*{Limitations}
Although our proposed Adaptive Stealing (AS) achieves higher attack performance compared to Watermark Stealing (WS) at a small number of victim watermarked texts (10,000), this performance improvement results in approximately 1.6× generation latency compared to WS. 
Furthermore, due to time and resource constraints in our experiments, we evaluate only two victim models (OPT-2.7b \& Llama3-8b) and three watermarks (KGW, SynthID, Unbiased). 
The multilingual attack effectiveness of AS remains to be tested.
Currently, we only focus on the English dataset scenario in this paper. 
Furthermore, AS applies the impression in the same manner as WS. 
While this approach effectively achieves the attack, it causes relatively large damage to text quality. 
Subsequent work could consider how to optimize the stealing watermark algorithm by improving the impression application method. 

\section*{Acknowledgement}
This work was supported in part by the National Natural Science Foundation of China under Grants 62372058, U22A2026.


\bibliography{acl_latex}

\begin{thebibliography}{44}
\providecommand{\natexlab}[1]{#1}

\bibitem[{AI@Meta(2024)}]{MO-l3-llama3modelcard}
AI@Meta. 2024.
\newblock \href {https://github.com/meta-llama/llama3/blob/main/MODEL_CARD.md}
  {Llama 3 model card}.

\bibitem[{Chang et~al.(2025)Chang, Hassani, and
  Shokri}]{AT-SC-Smoothing:DBLP:conf/emnlp/ChangHS25}
Hongyan Chang, Hamed Hassani, and Reza Shokri. 2025.
\newblock \href {https://aclanthology.org/2025.findings-emnlp.264/} {Watermark
  smoothing attacks against language models}.
\newblock In \emph{Findings of the Association for Computational Linguistics:
  {EMNLP} 2025, Suzhou, China, November 4-9, 2025}, pages 4915--4941.
  Association for Computational Linguistics.

\bibitem[{Chen and Shu(2024)}]{Harm-misinformation-DBLP:conf/iclr/ChenS24}
Canyu Chen and Kai Shu. 2024.
\newblock \href {https://openreview.net/forum?id=ccxD4mtkTU} {Can llm-generated
  misinformation be detected?}
\newblock In \emph{The Twelfth International Conference on Learning
  Representations, {ICLR} 2024, Vienna, Austria, May 7-11, 2024}.
  OpenReview.net.

\bibitem[{Chen et~al.(2025{\natexlab{a}})Chen, Wu, Guo, and
  Huang}]{AT-SP-Demark:DBLP:conf/icml/ChenWGH25}
Ruibo Chen, Yihan Wu, Junfeng Guo, and Heng Huang. 2025{\natexlab{a}}.
\newblock \href {https://proceedings.mlr.press/v267/chen25bq.html} {De-mark:
  Watermark removal in large language models}.
\newblock In \emph{Forty-second International Conference on Machine Learning,
  {ICML} 2025, Vancouver, BC, Canada, July 13-19, 2025}, Proceedings of Machine
  Learning Research. {PMLR} / OpenReview.net.

\bibitem[{Chen et~al.(2025{\natexlab{b}})Chen, Wu, Guo, and
  Huang}]{WM2-P-Ruibo-DBLP:journals/corr/abs-2502-11268}
Ruibo Chen, Yihan Wu, Junfeng Guo, and Heng Huang. 2025{\natexlab{b}}.
\newblock \href {https://doi.org/10.48550/ARXIV.2502.11268} {Improved unbiased
  watermark for large language models}.
\newblock \emph{CoRR}, abs/2502.11268.

\bibitem[{Christ et~al.(2024)Christ, Gunn, and
  Zamir}]{WM2-S-Christ-DBLP:conf/colt/ChristGZ24}
Miranda Christ, Sam Gunn, and Or~Zamir. 2024.
\newblock \href {https://proceedings.mlr.press/v247/christ24a.html}
  {Undetectable watermarks for language models}.
\newblock In \emph{The Thirty Seventh Annual Conference on Learning Theory,
  June 30 - July 3, 2023, Edmonton, Canada}, volume 247 of \emph{Proceedings of
  Machine Learning Research}, pages 1125--1139. {PMLR}.

\bibitem[{Conover et~al.(2023)Conover, Hayes, Mathur, Xie, Wan, Shah, Ghodsi,
  Wendell, Zaharia, and Xin}]{DS-Dolly:DatabricksBlog2023DollyV2}
Mike Conover, Matt Hayes, Ankit Mathur, Jianwei Xie, Jun Wan, Sam Shah, Ali
  Ghodsi, Patrick Wendell, Matei Zaharia, and Reynold Xin. 2023.
\newblock \href
  {https://www.databricks.com/blog/2023/04/12/dolly-first-open-commercially-viable-instruction-tuned-llm}
  {Free dolly: Introducing the world's first truly open instruction-tuned llm}.

\bibitem[{Dathathri et~al.(2024)Dathathri, See, Ghaisas, Huang, McAdam, Welbl,
  Bachani, Kaskasoli, Stanforth, Matejovicova
  et~al.}]{WM2-S-SynthID-dathathri2024scalable}
Sumanth Dathathri, Abigail See, Sumedh Ghaisas, Po-Sen Huang, Rob McAdam,
  Johannes Welbl, Vandana Bachani, Alex Kaskasoli, Robert Stanforth, Tatiana
  Matejovicova, and 1 others. 2024.
\newblock \href {https://www.nature.com/articles/s41586-024-08025-4} {Scalable
  watermarking for identifying large language model outputs}.
\newblock \emph{Nature}, 634(8035):818--823.

\bibitem[{Feng et~al.(2025)Feng, Zhang, Zhang, Zhang, and
  Pan}]{WM2-P-BiMark-DBLP:conf/icml/Feng0ZZP25}
Xiaoyan Feng, He~Zhang, Yanjun Zhang, Leo~Yu Zhang, and Shirui Pan. 2025.
\newblock \href {https://openreview.net/forum?id=Zvyb3WAg03} {Bimark: Unbiased
  multilayer watermarking for large language models}.
\newblock In \emph{Forty-second International Conference on Machine Learning,
  {ICML} 2025, Vancouver, BC, Canada, July 13-19, 2025}. OpenReview.net.

\bibitem[{Fraser et~al.(2025)Fraser, Dawkins, and
  Kiritchenko}]{Survey-Kath-DBLP:journals/jair/FraserDK25}
Kathleen~C. Fraser, Hillary Dawkins, and Svetlana Kiritchenko. 2025.
\newblock \href {https://doi.org/10.1613/JAIR.1.16665} {Detecting ai-generated
  text: Factors influencing detectability with current methods}.
\newblock \emph{J. Artif. Intell. Res.}, 82:2233--2278.

\bibitem[{Fu et~al.(2024)Fu, Xiong, and Dong}]{WM2-L-SW-DBLP:conf/aaai/FuXD24}
Yu~Fu, Deyi Xiong, and Yue Dong. 2024.
\newblock \href {https://doi.org/10.1609/AAAI.V38I16.29756} {Watermarking
  conditional text generation for {AI} detection: Unveiling challenges and a
  semantic-aware watermark remedy}.
\newblock In \emph{Thirty-Eighth {AAAI} Conference on Artificial Intelligence,
  {AAAI} 2024, Thirty-Sixth Conference on Innovative Applications of Artificial
  Intelligence, {IAAI} 2024, Fourteenth Symposium on Educational Advances in
  Artificial Intelligence, {EAAI} 2014, February 20-27, 2024, Vancouver,
  Canada}, pages 18003--18011. {AAAI} Press.

\bibitem[{Gu et~al.(2024)Gu, Li, Liang, and
  Hashimoto}]{AT-SP-learnable-DBLP:conf/iclr/GuLLH24}
Chenchen Gu, Xiang~Lisa Li, Percy Liang, and Tatsunori Hashimoto. 2024.
\newblock \href {https://openreview.net/forum?id=9k0krNzvlV} {On the
  learnability of watermarks for language models}.
\newblock In \emph{The Twelfth International Conference on Learning
  Representations, {ICLR} 2024, Vienna, Austria, May 7-11, 2024}.
  OpenReview.net.

\bibitem[{Hazell(2023)}]{Harm-autoPshing-DBLP:journals/corr/abs-2305-06972}
Julian Hazell. 2023.
\newblock \href {https://doi.org/10.48550/ARXIV.2305.06972} {Spear phishing
  with large language models}.
\newblock \emph{CoRR}, abs/2305.06972.

\bibitem[{He et~al.(2022{\natexlab{a}})He, Xu, Lyu, Wu, and
  Wang}]{WM3-Lexical-DBLP:conf/aaai/HeXLWW22}
Xuanli He, Qiongkai Xu, Lingjuan Lyu, Fangzhao Wu, and Chenguang Wang.
  2022{\natexlab{a}}.
\newblock \href {https://doi.org/10.1609/AAAI.V36I10.21321} {Protecting
  intellectual property of language generation apis with lexical watermark}.
\newblock In \emph{Thirty-Sixth {AAAI} Conference on Artificial Intelligence,
  {AAAI} 2022, Thirty-Fourth Conference on Innovative Applications of
  Artificial Intelligence, {IAAI} 2022, The Twelveth Symposium on Educational
  Advances in Artificial Intelligence, {EAAI} 2022 Virtual Event, February 22 -
  March 1, 2022}, pages 10758--10766. {AAAI} Press.

\bibitem[{He et~al.(2022{\natexlab{b}})He, Xu, Zeng, Lyu, Wu, Li, and
  Jia}]{WM3-Cater-DBLP:conf/nips/HeXZLWLJ22}
Xuanli He, Qiongkai Xu, Yi~Zeng, Lingjuan Lyu, Fangzhao Wu, Jiwei Li, and Ruoxi
  Jia. 2022{\natexlab{b}}.
\newblock \href
  {http://papers.nips.cc/paper\_files/paper/2022/hash/2433fec2144ccf5fea1c9c5ebdbc3924-Abstract-Conference.html}
  {{CATER:} intellectual property protection on text generation apis via
  conditional watermarks}.
\newblock In \emph{Advances in Neural Information Processing Systems 35: Annual
  Conference on Neural Information Processing Systems 2022, NeurIPS 2022, New
  Orleans, LA, USA, November 28 - December 9, 2022}.

\bibitem[{He et~al.(2024)He, Zhou, Hao, Liu, Wang, Tu, Zhang, and
  Wang}]{WM2-L-XSIR-DBLP:conf/acl/0002ZHL0T0W24}
Zhiwei He, Binglin Zhou, Hongkun Hao, Aiwei Liu, Xing Wang, Zhaopeng Tu,
  Zhuosheng Zhang, and Rui Wang. 2024.
\newblock \href {https://doi.org/10.18653/V1/2024.ACL-LONG.226} {Can watermarks
  survive translation? on the cross-lingual consistency of text watermark for
  large language models}.
\newblock In \emph{Proceedings of the 62nd Annual Meeting of the Association
  for Computational Linguistics (Volume 1: Long Papers), {ACL} 2024, Bangkok,
  Thailand, August 11-16, 2024}, pages 4115--4129. Association for
  Computational Linguistics.

\bibitem[{Hu et~al.(2024)Hu, Chen, Wu, Wu, Zhang, and
  Huang}]{WM2-P-UW-DBLP:conf/iclr/HuCWWZH24}
Zhengmian Hu, Lichang Chen, Xidong Wu, Yihan Wu, Hongyang Zhang, and Heng
  Huang. 2024.
\newblock \href {https://openreview.net/forum?id=uWVC5FVidc} {Unbiased
  watermark for large language models}.
\newblock In \emph{The Twelfth International Conference on Learning
  Representations, {ICLR} 2024, Vienna, Austria, May 7-11, 2024}.
  OpenReview.net.

\bibitem[{Huang et~al.(2025)Huang, Chen, and
  Shu}]{Harm-Authorship:10.1145/3715073.3715076}
Baixiang Huang, Canyu Chen, and Kai Shu. 2025.
\newblock \href {https://doi.org/10.1145/3715073.3715076} {Authorship
  attribution in the era of llms: Problems, methodologies, and challenges}.
\newblock \emph{SIGKDD Explor. Newsl.}, 26(2):21–43.

\bibitem[{Jovanovic et~al.(2024)Jovanovic, Staab, and
  Vechev}]{AT-SP-stealing-DBLP:conf/icml/0001SV24}
Nikola Jovanovic, Robin Staab, and Martin~T. Vechev. 2024.
\newblock \href {https://openreview.net/forum?id=Wp054bnPq9} {Watermark
  stealing in large language models}.
\newblock In \emph{Forty-first International Conference on Machine Learning,
  {ICML} 2024, Vienna, Austria, July 21-27, 2024}. OpenReview.net.

\bibitem[{Kirchenbauer et~al.(2023)Kirchenbauer, Geiping, Wen, Katz, Miers, and
  Goldstein}]{WM2-L-KGW-kirchenbauer2023watermark}
John Kirchenbauer, Jonas Geiping, Yuxin Wen, Jonathan Katz, Ian Miers, and Tom
  Goldstein. 2023.
\newblock \href {https://proceedings.mlr.press/v202/kirchenbauer23a.html} {A
  watermark for large language models}.
\newblock In \emph{International Conference on Machine Learning, {ICML} 2023,
  23-29 July 2023, Honolulu, Hawaii, {USA}}, volume 202 of \emph{Proceedings of
  Machine Learning Research}, pages 17061--17084. {PMLR}.

\bibitem[{Krishna et~al.(2023)Krishna, Song, Karpinska, Wieting, and
  Iyyer}]{AT-SC-DIPPER-DBLP:conf/nips/KrishnaSKWI23}
Kalpesh Krishna, Yixiao Song, Marzena Karpinska, John Wieting, and Mohit Iyyer.
  2023.
\newblock \href
  {http://papers.nips.cc/paper\_files/paper/2023/hash/575c450013d0e99e4b0ecf82bd1afaa4-Abstract-Conference.html}
  {Paraphrasing evades detectors of ai-generated text, but retrieval is an
  effective defense}.
\newblock In \emph{Advances in Neural Information Processing Systems 36: Annual
  Conference on Neural Information Processing Systems 2023, NeurIPS 2023, New
  Orleans, LA, USA, December 10 - 16, 2023}.

\bibitem[{Kuditipudi et~al.(2024)Kuditipudi, Thickstun, Hashimoto, and
  Liang}]{WM2-S-EXP-DBLP:journals/tmlr/KuditipudiTHL24}
Rohith Kuditipudi, John Thickstun, Tatsunori Hashimoto, and Percy Liang. 2024.
\newblock \href {https://openreview.net/forum?id=FpaCL1MO2C} {Robust
  distortion-free watermarks for language models}.
\newblock \emph{Trans. Mach. Learn. Res.}, 2024.

\bibitem[{Labadze et~al.(2023)Labadze, Grigolia, and
  Machaidze}]{Harm-academicfraud:labadze2023role}
Lasha Labadze, Maya Grigolia, and Lela Machaidze. 2023.
\newblock \href {https://link.springer.com/article/10.1186/s41239-023-00426-1}
  {Role of ai chatbots in education: systematic literature review}.
\newblock \emph{International journal of Educational Technology in Higher
  education}, 20(1):56.

\bibitem[{Liu et~al.(2024)Liu, Pan, Hu, Meng, and
  Wen}]{WM2-L-SIR-DBLP:conf/iclr/LiuPHM024}
Aiwei Liu, Leyi Pan, Xuming Hu, Shiao Meng, and Lijie Wen. 2024.
\newblock \href {https://openreview.net/forum?id=6p8lpe4MNf} {A semantic
  invariant robust watermark for large language models}.
\newblock In \emph{The Twelfth International Conference on Learning
  Representations, {ICLR} 2024, Vienna, Austria, May 7-11, 2024}.
  OpenReview.net.

\bibitem[{Liu et~al.(2025)Liu, Pan, Lu, Li, Hu, Zhang, Wen, King, Xiong, and
  Yu}]{Survey-liu-DBLP:journals/csur/LiuPLLHZWKXY25}
Aiwei Liu, Leyi Pan, Yijian Lu, Jingjing Li, Xuming Hu, Xi~Zhang, Lijie Wen,
  Irwin King, Hui Xiong, and Philip~S. Yu. 2025.
\newblock \href {https://doi.org/10.1145/3691626} {A survey of text
  watermarking in the era of large language models}.
\newblock \emph{{ACM} Comput. Surv.}, 57(2):47:1--47:36.

\bibitem[{Lu et~al.(2024)Lu, Liu, Yu, Li, and
  King}]{WM2-L-EWD-DBLP:conf/acl/LuLY0K24}
Yijian Lu, Aiwei Liu, Dianzhi Yu, Jingjing Li, and Irwin King. 2024.
\newblock \href {https://doi.org/10.18653/V1/2024.ACL-LONG.630} {An
  entropy-based text watermarking detection method}.
\newblock In \emph{Proceedings of the 62nd Annual Meeting of the Association
  for Computational Linguistics (Volume 1: Long Papers), {ACL} 2024, Bangkok,
  Thailand, August 11-16, 2024}, pages 11724--11735. Association for
  Computational Linguistics.

\bibitem[{OpenAI(2023)}]{MO-GPT4:DBLP:journals/corr/abs-2303-08774}
OpenAI. 2023.
\newblock \href {https://doi.org/10.48550/ARXIV.2303.08774} {{GPT-4} technical
  report}.
\newblock \emph{CoRR}, abs/2303.08774.

\bibitem[{Pan et~al.(2024)Pan, Liu, He, Gao, Zhao, Lu, Zhou, Liu, Hu, Wen,
  King, and Yu}]{TOOL-Markllm-DBLP:journals/corr/abs-2405-10051}
Leyi Pan, Aiwei Liu, Zhiwei He, Zitian Gao, Xuandong Zhao, Yijian Lu, Binglin
  Zhou, Shuliang Liu, Xuming Hu, Lijie Wen, Irwin King, and Philip~S. Yu. 2024.
\newblock \href {https://doi.org/10.18653/v1/2024.emnlp-demo.7} {{M}ark{LLM}:
  An open-source toolkit for {LLM} watermarking}.
\newblock In \emph{Proceedings of the 2024 Conference on Empirical Methods in
  Natural Language Processing: System Demonstrations}, pages 61--71, Miami,
  Florida, USA. Association for Computational Linguistics.

\bibitem[{Pan et~al.(2025)Pan, Liu, Huang, Lu, Hu, Wen, King, and
  Yu}]{AT-SP-Training-DBLP:conf/acl/PanL0LH0KY25}
Leyi Pan, Aiwei Liu, Shiyu Huang, Yijian Lu, Xuming Hu, Lijie Wen, Irwin King,
  and Philip~S. Yu. 2025.
\newblock \href {https://aclanthology.org/2025.acl-long.648/} {Can {LLM}
  watermarks robustly prevent unauthorized knowledge distillation?}
\newblock In \emph{Proceedings of the 63rd Annual Meeting of the Association
  for Computational Linguistics (Volume 1: Long Papers), {ACL} 2025, Vienna,
  Austria, July 27 - August 1, 2025}, pages 13228--13251. Association for
  Computational Linguistics.

\bibitem[{Peng et~al.(2023)Peng, Yi, Wu, Wu, Zhu, Lyu, Jiao, Xu, Sun, and
  Xie}]{WM1-EaaS-DBLP:conf/acl/PengYWWZLJXSX23}
Wenjun Peng, Jingwei Yi, Fangzhao Wu, Shangxi Wu, Bin Zhu, Lingjuan Lyu,
  Binxing Jiao, Tong Xu, Guangzhong Sun, and Xing Xie. 2023.
\newblock \href {https://doi.org/10.18653/V1/2023.ACL-LONG.423} {Are you
  copying my model? protecting the copyright of large language models for eaas
  via backdoor watermark}.
\newblock In \emph{ACL}, pages 7653--7668.

\bibitem[{Raffel et~al.(2020)Raffel, Shazeer
  et~al.}]{DS-C4:journals/jmlr/RaffelSRLNMZLL20}
Colin Raffel, Noam Shazeer, and 1 others. 2020.
\newblock \href {http://jmlr.org/papers/v21/20-074.html} {Exploring the limits
  of transfer learning with a unified text-to-text transformer}.
\newblock \emph{J. Mach. Learn. Res.}, 21:140:1--140:67.

\bibitem[{Sadasivan et~al.(2023)Sadasivan, Kumar, Balasubramanian, Wang, and
  Feizi}]{AT-SP-181cww-DBLP:journals/corr/abs-2303-11156}
Vinu~Sankar Sadasivan, Aounon Kumar, Sriram Balasubramanian, Wenxiao Wang, and
  Soheil Feizi. 2023.
\newblock \href {https://doi.org/10.48550/ARXIV.2303.11156} {Can ai-generated
  text be reliably detected?}
\newblock \emph{CoRR}, abs/2303.11156.

\bibitem[{Shaikh et~al.(2023)Shaikh, Zhang, Held, Bernstein, and
  Yang}]{DS-harmfulq:DBLP:conf/acl/Shaikh0HBY23}
Omar Shaikh, Hongxin Zhang, William Held, Michael~S. Bernstein, and Diyi Yang.
  2023.
\newblock \href {https://doi.org/10.18653/V1/2023.ACL-LONG.244} {On second
  thought, let's not think step by step! bias and toxicity in zero-shot
  reasoning}.
\newblock In \emph{Proceedings of the 61st Annual Meeting of the Association
  for Computational Linguistics (Volume 1: Long Papers), {ACL} 2023, Toronto,
  Canada, July 9-14, 2023}, pages 4454--4470. Association for Computational
  Linguistics.

\bibitem[{Team(2024)}]{MO-gem2:gemma_2024}
Gemma Team. 2024.
\newblock \href {https://doi.org/10.34740/KAGGLE/M/3301} {Gemma}.

\bibitem[{Wang and Li(2025)}]{Harm-tracingLLM-Wang2025OnCT}
Quan Wang and Haoran Li. 2025.
\newblock \href {https://api.semanticscholar.org/CorpusID:276543233} {On
  continually tracing origins of llm-generated text and its application in
  detecting cheating in student coursework}.
\newblock \emph{Big Data Cogn. Comput.}, 9:50.

\bibitem[{Wong et~al.(2025)Wong, Zhou, Zhou, and
  Si}]{WM2-L-E2E-DBLP:conf/icml/WongZ0S25}
Kahim Wong, Jicheng Zhou, Jiantao Zhou, and Yain{-}Whar Si. 2025.
\newblock \href {https://openreview.net/forum?id=9sNiCqi2RD} {An end-to-end
  model for logits-based large language models watermarking}.
\newblock In \emph{Forty-second International Conference on Machine Learning,
  {ICML} 2025, Vancouver, BC, Canada, July 13-19, 2025}. OpenReview.net.

\bibitem[{Xu et~al.(2024)Xu, Yao, and
  Liu}]{WM1-RL-DBLP:journals/corr/abs-2403-10553}
Xiaojun Xu, Yuanshun Yao, and Yang Liu. 2024.
\newblock \href {https://doi.org/10.48550/ARXIV.2403.10553} {Learning to
  watermark llm-generated text via reinforcement learning}.
\newblock \emph{CoRR}, abs/2403.10553.

\bibitem[{Yang et~al.(2024)Yang, Yang, Hui, Zheng, Yu, Zhou, Li, Li, Liu,
  Huang, Dong, Wei, Lin, Tang, Wang, Yang, Tu, Zhang, Ma, Xu, Zhou, Bai, He,
  Lin, Dang, Lu, Chen, Yang, Li, Xue, Ni, Zhang, Wang, Peng, Men, Gao, Lin,
  Wang, Bai, Tan, Zhu, Li, Liu, Ge, Deng, Zhou, Ren, Zhang, Wei, Ren, Fan, Yao,
  Zhang, Wan, Chu, Liu, Cui, Zhang, and Fan}]{MO-Qwen:qwen2}
An~Yang, Baosong Yang, Binyuan Hui, Bo~Zheng, Bowen Yu, Chang Zhou, Chengpeng
  Li, Chengyuan Li, Dayiheng Liu, Fei Huang, Guanting Dong, Haoran Wei, Huan
  Lin, Jialong Tang, Jialin Wang, Jian Yang, Jianhong Tu, Jianwei Zhang,
  Jianxin Ma, and 40 others. 2024.
\newblock \href {https://arxiv.org/abs/2412.15115} {Qwen2 technical report}.
\newblock \emph{arXiv preprint arXiv:2407.10671}.

\bibitem[{Yang et~al.(2022)Yang, Zhang, Chen, Zhang, Ma, Wang, and
  Yu}]{WM3-Lexical22}
Xi~Yang, Jie Zhang, Kejiang Chen, Weiming Zhang, Zehua Ma, Feng Wang, and
  Nenghai Yu. 2022.
\newblock \href {https://doi.org/10.1609/AAAI.V36I10.21415} {Tracing text
  provenance via context-aware lexical substitution}.
\newblock In \emph{AAAI}, pages 11613--11621.

\bibitem[{Zhang et~al.(2023)Zhang, Edelman, Francati, Venturi, Ateniese, and
  Barak}]{AT-SC-sand-DBLP:journals/iacr/ZhangEFVAB23}
Hanlin Zhang, Benjamin~L. Edelman, Danilo Francati, Daniele Venturi, Giuseppe
  Ateniese, and Boaz Barak. 2023.
\newblock \href {https://eprint.iacr.org/2023/1776} {Watermarks in the sand:
  Impossibility of strong watermarking for generative models}.
\newblock \emph{{IACR} Cryptol. ePrint Arch.}, page 1776.

\bibitem[{Zhang et~al.(2022)Zhang, Roller et~al.}]{MO-opt27-zhang2022opt}
Susan Zhang, Stephen Roller, and 1 others. 2022.
\newblock \href {https://doi.org/10.48550/ARXIV.2205.01068} {{OPT:} open
  pre-trained transformer language models}.
\newblock \emph{CoRR}, abs/2205.01068.

\bibitem[{Zhang et~al.(2024)Zhang, Zhang, Zhang, Zhang, Chen, Hu, Gill, and
  Pan}]{AT-SP-greenlistPredict-DBLP:conf/acsac/ZhangZZZ0HGP24}
Zhaoxi Zhang, Xiaomei Zhang, Yanjun Zhang, Leo~Yu Zhang, Chao Chen, Shengshan
  Hu, Asif Gill, and Shirui Pan. 2024.
\newblock \href {https://doi.org/10.1109/ACSAC63791.2024.00021} {Stealing
  watermarks of large language models via mixed integer programming}.
\newblock In \emph{Annual Computer Security Applications Conference, {ACSAC}
  2024, Honolulu, HI, USA, December 9-13, 2024}, pages 46--60. {IEEE}.

\bibitem[{Zhao et~al.(2024)Zhao, Ananth, Li, and
  Wang}]{WM2-L-Unigram-DBLP:conf/iclr/ZhaoA0W24}
Xuandong Zhao, Prabhanjan~Vijendra Ananth, Lei Li, and Yu{-}Xiang Wang. 2024.
\newblock \href {https://openreview.net/forum?id=SsmT8aO45L} {Provable robust
  watermarking for ai-generated text}.
\newblock In \emph{The Twelfth International Conference on Learning
  Representations, {ICLR} 2024, Vienna, Austria, May 7-11, 2024}.
  OpenReview.net.

\bibitem[{Zou et~al.(2023)Zou, Wang, Kolter, and
  Fredrikson}]{DS-advbench:DBLP:journals/corr/abs-2307-15043}
Andy Zou, Zifan Wang, J.~Zico Kolter, and Matt Fredrikson. 2023.
\newblock \href {https://doi.org/10.48550/ARXIV.2307.15043} {Universal and
  transferable adversarial attacks on aligned language models}.
\newblock \emph{CoRR}, abs/2307.15043.

\end{thebibliography}

\appendix

\section{Algorithms}
\label{sec:app-algs}
In this section, we introduce the WS algorithm \cite{AT-SP-stealing-DBLP:conf/icml/0001SV24} and victim watermark algorithms. 
We also detail the parameters of each watermark. 
All our watermarks are implemented using Markllm \cite{TOOL-Markllm-DBLP:journals/corr/abs-2405-10051}. 

\subsection{KGW}
The beginning design of KGW \cite{WM2-L-KGW-kirchenbauer2023watermark} is LeftHash-scheme. 
LeftHash-scheme specifies that $Seal^\theta(\cdot)$ only utilize the most left token in $ctx$. 
Additionally, other Hash-schemes exist, such as MinHash-scheme which utilizes the token with the minimum hash value in $ctx$, and MaxHash-scheme which conversely utilizes the token with the maximum hash value. 
Initially, KGW sets $|ctx|=1$.
When generating $T_{h+1}$, $Seal^\theta(\cdot)$ inputs the last token $T_{h}$ and the secret key $\mathcal{K}$. 
At this time, $Seal^\theta(\cdot)$ generates a code $c$ which is a random number seed, deterministically partitioning the vocabulary into red $\mathcal{R}$ and green $\mathcal{G}$ based on the random seed $c$. 
Among them, the proportion of green tokens is $\gamma$. 
Then KGW constructs a vector as its impression, which has values that green tokens are $\delta$, red tokens is 0. 
Following standard KGW implementation, we set $\delta=2$.
The vector is the impression $im$. 
Subsequently, KGW increases the generation probability of green tokens through the following formula:
\begin{equation}
    \label{eq:app-kgw} 
    \begin{aligned}
        &\hat{p}^{h+1}_{T}=
        \begin{cases}
            \frac{exp(l^{h+1}_T + \delta )}{ L_{sum}}, &T \in \mathcal{G} \\
            \frac{exp(l^{h+1}_T)}{ L_{sum}}, &T \in \mathcal{R}
        \end{cases} \\
        &L_{sum}=\sum_{j\in \mathcal{R}} exp(l_j^{h+1} ) + \sum_{j \in \mathcal{G}} exp(l_j^{h+1} + \delta)
    \end{aligned}
\end{equation}
$\delta$ is a positive constant, $L_{sum}$ represents the sum of modified logits. 
$\hat{p}^{h+1}_{T}$ represents the probability of $T$ after watermarking at $h+1$ step.
 
At detection stage, KGW can obtain the value of $c$ at each step through $Seal^\theta(\cdot)$, and then determine whether each token $T$ in the text is green. 
KGW counts the number of green tokens in the text as $n_g$. 
Through z-statistic $z=(n_g-\gamma L)/\sqrt{L \gamma(1-\gamma)}$, KGW obtains $z$ to represents the confidence that the text is watermarked. 
In this paper, $z$ is equivalent to $WCS$. 

We set $\gamma=0.5$ and LeftHash-scheme for KGW in this paper, and provide our Markllm-style parameter files in the future.

\subsection{SelfHash-Scheme in KGW}
SelfHash-scheme KGW incorporates $T_{h+1}$, the generating token at this time, to participate in $Seal^{\theta}(\cdot)$. 
When constructing red and green lists using $c$ generated by $T_{h+1}$, the generation process ensures $T_{h+1}$ belongs to the green list determined by its own hash value.
And the method to construct a impression $im$ is presented in the following equation:
\begin{equation}
\begin{aligned}
    im=\min\{H(T_{h-|ctx|+1}),...H(T_{h}),  \\  H(T_{h+1})\} \cdot\mathcal{K} \cdot H(T_{h+1})
\end{aligned}
\end{equation}
Subsequent operations remain consistent with KGW. 

SelfHash-scheme KGW is the watermark specifically targeted by WS. 
In WS, the author sets $|ctx|=2$ for SelfHash-scheme KGW. 
Although SelfHash-scheme KGW is a high-security-performance watermark, its high resource consumption renders it impractical for real-world deployment. 

In our experiments, text generation with OPT-2.7b using SelfHash-scheme KGW requires \textbf{over 300 seconds} for 200-token generation in a single  Tesla-V100 GPU, while LeftHash-scheme averages \textbf{below 10 seconds} (measured in our Markllm implementation). 
Therefore, considering the efficiency factor, the SelfHash-scheme KGW is not evaluated in this paper. 

\subsection{Watermark Stealing (WS)}
WS construct three seals and statically weighted them to obtain the final seal. 

The first seal focuses on the overall $ctx$ information. 
WS utilize transformation function $H_1(\cdot)$ to convert $ctx$ to a token set $k^{H_1}$. 
WS defines  $\{\cdot\}$ as the token set representation. 
Given $ctx=[T_1,T_2,T_3]$, it derives $k^{H_1}=\{T_1,T_2,T_3\}$. 
$k^{H_1}$ loses the ordered information of $ctx$, which is the largest difference compared to the function $H^o(ctx,|ctx|-1)$ in AS. 
Through Eq. (\ref{eq:as-score-function}), WS obtains the impression $s^{H_1}$ corresponding to $H_1(\cdot)$ based on the relevant $ctx$. 
The impression is derived through concatenation of token scores from the whole vocabulary, similar to the AS process. 

The second seal is more complex. 
To construct it, WS obtains two format impressions. 
WS lets $s_i$ and $s_{ij}$ denotes the impressions $\mathcal{S}(\cdot, \{T_i\}])$ and $\mathcal{S}(\cdot, \{T_i,T_j\}])$
By defining $cossim(\cdot)$ as the function for cosine similarity, WS searches a unique i, s.t.
\begin{equation}
    cossim(s_i,s_{ij})>cossim(s_j,s_{ij}), \forall j \neq i
\end{equation}
The transformation function $H_2(\cdot)$ select this token $T_i$ as $k^{H_2}$. 
Subsequently, WS constructs the corresponding impression $s^{H_2}$ according to $\mathcal{S}(T,k^{H_2})$. 

The transformation function $H_3(\cdot)$ of the third seal transforms all possible $ctx$ into an empty set $k^{H_3}=\{\}$, which ignores the information from $ctx$. 
Its corresponding score for $T$ is $\mathcal{S}(T,\{\})$ and the impression is $s^{H_3}$. 

The final impression is formulated as:
\begin{equation}
    \hat{s}=\frac{1}{w_1+w_2+w_3} (w_1 \cdot s^{H_1}+w_2 \cdot s^{H_2} + w_3 \cdot s^{H_3})
    \label{eq:app-steal-weight-score}
\end{equation}

The second seal is specifically designed by WS authors to target SelfHash-scheme KGW, and ablation studies conducted in WS's work demonstrate its effectiveness. 
However, our experimental results reveal that introducing the second seal adversely affects the stealing outcomes of certain watermarks (Table \ref{tab:app-steal-ch}). 
Meanwhile, we find that solely adopting the first seal can comprehensively address diverse watermarks, which is the configuration adopted in this paper $(w_1,w_2,w_3)=(1,0,0)$. 
The experiments will be demonstrated in subsequent sections.

\subsection{SynthID} 
During the context code generation phase, SynthID utilizes hash functions to combine $ctx$ with multiple predefined secret keys sequentially, generating several distinct integers as random number seeds.
The ordered sequence of these random number seeds constitutes the random sequence $rs$. 
Subsequently, SynthID sequentially constructs $m$ random number generators based on the seeds in $rs$. 
The $m$ random number generators are $g_1,...g_m$, and each generator $g$ assigns binary scores (0/1) to all tokens in vocabulary $V$. 
By concatenating the scores of all tokens, we can obtain the impression. 

When generating $T_{h+1}$, SynthID samples $2^m$ tokens from $p^{h+1}$ and divides them into pairs of competing tokens. 
At this stage, $g_1$ evaluates paired tokens and selects high-score tokens for the next round. 
In the second round, $g_2$ evaluates paired tokens and selects high-score tokens. 
SynthID executes this iterative cycle until $g_m$ selects the final winning token. 
The final winning token is selected as the result token $\hat{T}_{h+1}$ in SynthID. 

During detection, SynthID regenerates $rs$ for each token, enabling all $g$ to conduct scoring evaluations.
The watermark confidence score for token $T$ is calculated as $\sum_{i=1}^{m}g_i(T)$.
The mean value of scores across all tokens in text represents the watermark confidence of the text, which is $WCS$ in this paper. 

In the Markllm configuration, the parameter $m$ is set to 30. 
Markllm employs an acceleration method that avoids actual sampling of $2^m$ tokens to maintain watermarking efficiency. 

\subsection{Unbiased} 
After obtaining the code $c$ like KGW, utilizes the code $c$ as a seed to randomly permute the vocabulary into a token list $O=\{o_1, ..., o_{|V|}\}$.
Unbiased designs a reweighting method to transform $p^{h+1}$ using the cumulative distribution function concept, $F_i=\sum_{j=0}^i(p_{o_j})$.
The reweighting method is defined as follows: 
\begin{equation}
    \label{eq:Unbiased}
    p_{o_i}^{w}=\begin{cases}
        0,                  & F_i < \frac{1}{2} \\
        2(F_i-\frac{1}{2}),                    & F_i \geq \frac{1}{2}, F_{i-1} < \frac{1}{2} \\
        2(F_i - F_{i-1})    & F_{i-1}\geq \frac{1}{2}
    \end{cases}
\end{equation}
Then Unbiased transform $p^{h+1}$ by Eq. (\ref{eq:Unbiased}) to $\hat{p}^{h+1}$. 

During detection, Unbiased accumulates the watermark probabilities $\hat{p}^{h+1}$ of each token $T$ in the text as a detection metric, which is $WCS$ in this paper.

\section{Experimental Settings} 
\label{sec:app-exp-set}
\subsection{Hardware Setting \& License}
All experiments are conducted on the Linux system using two Tesla-V100 GPUs. 
We use PyTorch 2.2.0 and Transformers library 4.45.2. 
And we utilize a tool, Markllm to assist our watermark implementation. 
The version of Markllm is 0.1.5. 

For license information, Our experiments utilize C4 dataset (ODC-By 1.0 license), Dolly and Harm datasets (from the open-source code os WS, both under MIT license). 
We use OPT-2.7b (OPT-175B license), Llama3-8b (META LLAMA 3 COMMUNITY license), Gemma2-2b (gemma license) and Qwen2.5-7b (Apache 2.0 license). The watermark implementations leverage Markllm toolkit (Apache 2.0 license). 
All datasets and models are used strictly for non-commercial research purposes consistent with their license terms. 
Upon acceptance, we will release our AS code under the Apache 2.0 license with explicit restrictions limiting its use to academic research and watermark robustness evaluation. 

\subsection{Preparation of \texorpdfstring{$D_w$}{D_w}}
\label{sec:app-expset-prepare-dw}
We set higher restrictions for the preparation of our watermarked victim text. 
For each text example in C4, we extract the first 30 tokens as a prompt and generate 400 new tokens. 
Given each watermark setting, we generate $10^4$ watermarked texts to construct $D_w$. 
A corresponding dataset $D_n$ of equal size containing non-watermarked texts is also prepared.

\subsection{Default Attack Parameters} 
\label{sec:app-expset-attack}
Following WS, we set the clipping parameter to $c=2$ in all experiments. 
Across all watermark configurations, we fix the context length to $|ctx|=3$ for attack execution.
About AS, we configure $k=128$ for Adaptive Selection. 
For spoofing attack, we set $\delta_{att} = 4$ and $LM_{att}$=Qwen2.5-7b \cite{MO-Qwen:qwen2}. 
In contrast, we define $\delta_{att} = -4$ and $LM_{att}$=Dipper \cite{AT-SC-DIPPER-DBLP:conf/nips/KrishnaSKWI23} for scrubbing attack.
Dipper is a current state-of-the-art paraphraser. 

Under the above conditions, our evaluation performed a single attack, obtaining the average at the dataset level as the final result.

\section{Supplementary Experiments}

\subsection{Adaptive Selection Analysis}
\label{sec:app-exp-abl-select}

\begin{table}[tbp]
\centering
\scalebox{0.7}{
\begin{tabular}{@{}ccccc@{}}
\toprule
        & \multicolumn{2}{c}{$LM_{vic}$=OPT-2.7b} & \multicolumn{2}{c}{$LM_{vic}$=Llama3-8b} \\
        & Dolly              & Harm              & Dolly              & Harm               \\ \midrule
w/o WC  & 2.454              & 2.710             & 2.424              & 2.607              \\
w/o GP  & 0.358              & 0.172             & 0.438              & 0.381              \\
w/o DGR & 2.666              & 2.752             & 2.615              & 2.725              \\
AS      & 2.698              & 2.770             & 2.646              & 2.648              \\ \bottomrule
\end{tabular}
}
\caption{Ablation study of Adaptive Selection components (WC, GP, DGR) for spoofing attacks. Victim watermark is KGW, evaluation metric is WCS.}
\label{tab:ablation-selection}
\end{table}
We perform an ablation study on three criteria of Adaptive Selection, with results shown in Table \ref{tab:ablation-selection}.

We find that GP plays a crucial role in Adaptive Selection, and WC also significantly improves Adaptive Selection's results.
In Table \ref{tab:ablation-selection}, AS's results are far superior to w/o GP, and also demonstrate a certain advantage over w/o WC.
However, DGR's advantage is not obvious, and even in $LM_{vic}$=Llama3-8b, w/o DGR achieves better performance when using the Harm dataset.

\begin{table}[tbp]
\centering
\scalebox{0.7}{
\begin{tabular}{@{}ccccc@{}}
\toprule
       & \multicolumn{2}{c}{$LM_{vic}$=OPT-2.7b}                               & \multicolumn{2}{c}{$LM_{vic}$=Llama3-8b} \\
       & Dolly                             & Harm                              & Dolly              & Harm                \\ \midrule
top32  & \phantom{0}0.020 & \phantom{0}0.097 & \phantom{0}0.018              & -0.069              \\
top64  & -0.020                            & \phantom{0}0.020                             & \phantom{0}0.011              & \phantom{0}0.032               \\
top128 & \phantom{0}0.020                             & \phantom{0}0.060                             & \phantom{0}0.024              & \phantom{0}0.016               \\
top256 & \phantom{0}0.021                             & \phantom{0}0.048                             & \phantom{0}0.003              & \phantom{0}0.044               \\
top512 & \phantom{0}0.031                             & \phantom{0}0.101                             & \phantom{0}0.017              & \phantom{0}0.012               \\ \bottomrule
\end{tabular}
}
\caption{The confrontation results of DGR and non-DGR settings.}
\label{tab:topk-dgr}
\end{table}
\begin{table*}[t!]
\centering
\scalebox{0.68}{
	
	\begin{tabular}{@{}cccccccccccccc@{}}
	\toprule
	                            &                   & \multicolumn{6}{c}{$LM_{vic}$=OPT-2.7b}               & \multicolumn{6}{c}{$LM_{vic}$=Llama3-8b}              \\ \cmidrule(lr){3-8} \cmidrule(lr){9-14}
	                            &                   & \multicolumn{3}{c}{Dolly} & \multicolumn{3}{c}{Harm} & \multicolumn{3}{c}{Dolly} & \multicolumn{3}{c}{Harm} \\ \cmidrule(lr){3-5} \cmidrule(lr){6-8} \cmidrule(lr){9-11} \cmidrule(lr){12-14}
	                        $LM_{att}$    & SWA              & WCS    & AUC   & TPR@1\%  & WCS    & AUC   & TPR@1\% & WCS    & AUC   & TPR@1\%  & WCS    & AUC   & TPR@1\% \\ \midrule
	\multirow{2}{*}{OPT-1.3b}   & WS                & 4.140  & 0.99  & 0.80     & 4.290  & 0.99  & 0.89    & 2.925  & 0.94  & 0.55     & 2.974  & 0.95  & 0.59    \\
	                            & \textbf{AS(Ours)} & 5.060  & 1.00  & 0.92     & 5.395  & 1.00  & 0.98    & 4.156  & 0.99  & 0.83     & 4.174  & 0.99  & 0.88    \\ \midrule
	\multirow{2}{*}{OPT-2.7b}   & WS                & 4.054  & 0.98  & 0.74     & 4.144  & 0.98  & 0.79    & 2.876  & 0.94  & 0.53     & 2.900  & 0.94  & 0.56    \\
	                            & \textbf{AS(Ours)} & 5.058  & 0.99  & 0.93     & 5.263  & 1.00  & 0.93    & 4.062  & 0.98  & 0.82     & 4.296  & 0.99  & 0.90    \\ \midrule
	\multirow{2}{*}{Gemma2-2b}  & WS                & 2.710  & 0.87  & 0.45     & 2.796  & 0.91  & 0.44    & 2.076  & 0.83  & 0.35     & 2.342  & 0.87  & 0.43    \\
	                            & \textbf{AS(Ours)} & 4.284  & 0.95  & 0.76     & 4.626  & 0.98  & 0.81    & 3.589  & 0.93  & 0.70     & 4.039  & 0.97  & 0.82    \\ \midrule
	\multirow{2}{*}{Llama3-8b}  & WS                & 2.613  & 0.89  & 0.38     & 2.872  & 0.92  & 0.46    & 3.086  & 0.93  & 0.59     & 3.364  & 0.95  & 0.67    \\
	                            & \textbf{AS(Ours)} & 4.455  & 0.98  & 0.81     & 4.529  & 0.98  & 0.83    & 4.696  & 0.99  & 0.90     & 5.001  & 0.98  & 0.93    \\ \midrule
	\multirow{2}{*}{Qwen2.5-7b} & WS                & 1.228  & 0.74  & 0.06     & 1.404  & 0.77  & 0.11    & 1.333  & 0.74  & 0.15     & 1.431  & 0.75  & 0.19    \\
	                            & \textbf{AS(Ours)} & 2.698  & 0.91  & 0.36     & 2.770  & 0.92  & 0.42    & 2.646  & 0.89  & 0.47     & 2.648  & 0.90  & 0.48    \\ \bottomrule
	\end{tabular}
}
\caption{Results of spoofing attacks on KGW with different $LM_{att}$.}
\label{tab:dif-model}
\end{table*}
We do not assert that DGR is ineffective. 
DGR itself is analogous to top-k sampling in LLM generation, while removing DGR is analogous to setting $p=1$ in top-p sampling. 
Therefore, DGR eliminates the interference of some low-probability tokens, which theoretically improves text quality without affecting detectability.
However, its effect is minimal, akin to the minimal difference between top-k and top-p sampling methods. 

To prove the effect of DGR, we establish a more fine-grained evaluation method. 
For each sample, we pit DGR and non-DGR settings against each other on detectability and text quality. 
When a sample wins on both detectability and text quality, record 1 point; loses on both, record -1 point; otherwise, record 0 points. 
We set different $k$ for DGR, and their average point results are shown in Table \ref{tab:topk-dgr}.
The experimental results show that although DGR does not always enhance the attack effectiveness, from an overall perspective, DGR has a promoting effect on AS results. 
Therefore, we advocate deploying DGR for AS. 

Theoretically, WC is decided by empirical estimation on $D_w$, while the advantage of DGR is analogous to top-K during generation. 
This indicates that both methods require a substantial volume of watermarked text $D_w$ to ensure their effectiveness, and may introduce perturbations when operating with limited sample sizes.
Our $D_w$ has only 10k samples, and its data demand is less compared to the current feasible SWAs \cite{AT-SP-stealing-DBLP:conf/icml/0001SV24,AT-SP-Training-DBLP:conf/acl/PanL0LH0KY25}. 
Although the performance advantages of WC and DGR are marginal in Table \ref{tab:ablation-selection}, they enhance the adversarial capability of the AS as the sample size increases, and do not interfere with the advantage of the AS under low-sample regimes.
In Figure \ref{fig:exp-learnnum}, at a sample size of 2k, the low-sample fluctuations observed in WC and DGR do not compromise the performance advantage of the AS. 

\subsection{\texorpdfstring{$LM_{att}$}{LM_{att}} Analysis}

\begin{figure}[tbp]
  \centering 
  \includegraphics[width=\linewidth]{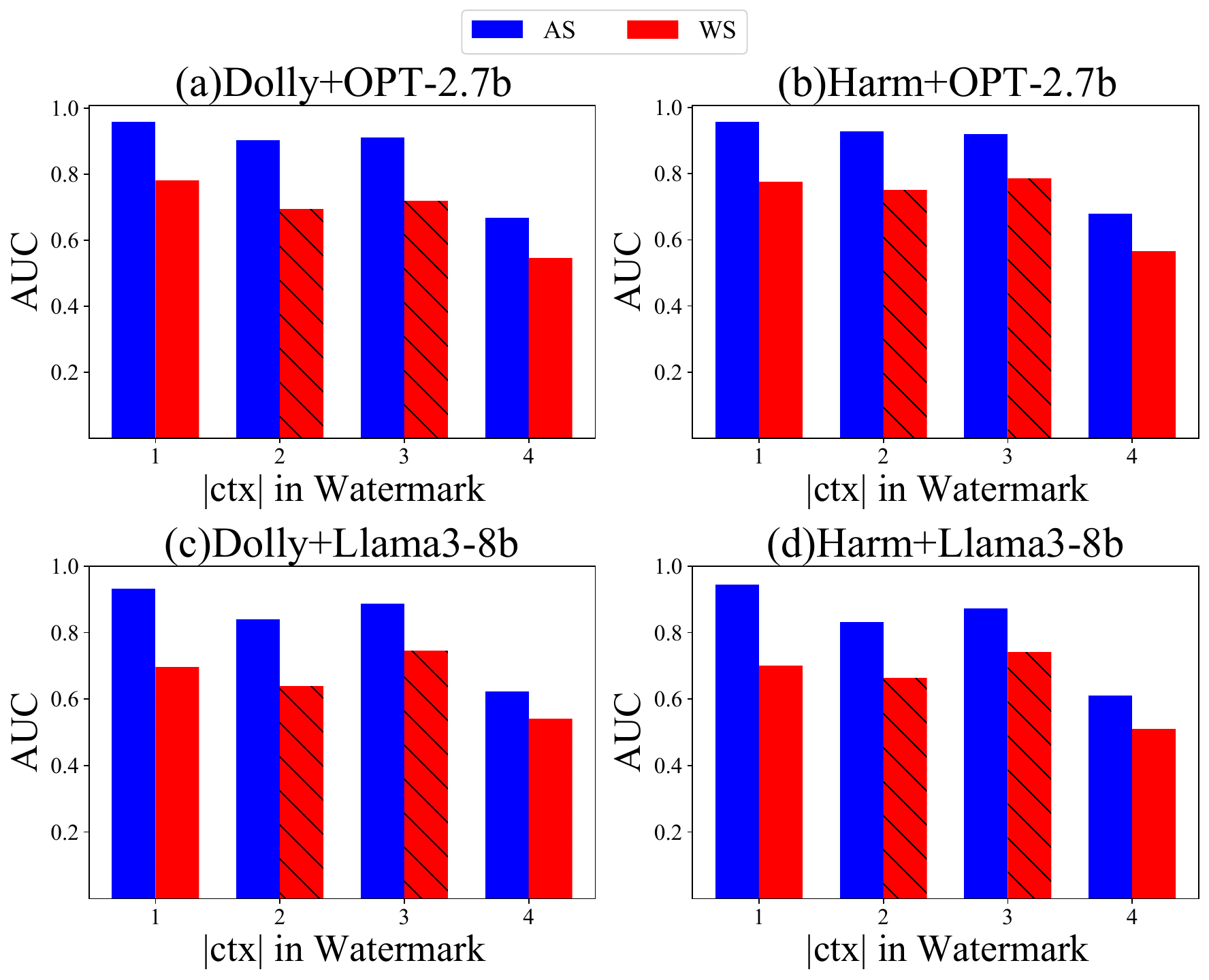}  
  \caption{Results of spoofing attacks on KGW with different lengths of $ctx$.}  
  \label{fig:exp-context-length}  
\end{figure}
We set five $LM_{att}$: OPT-1.3B and OPT-2.7b \cite{MO-opt27-zhang2022opt}, Gemma2-2B \cite{MO-gem2:gemma_2024}, Llama3-8b \cite{MO-l3-llama3modelcard} and Qwen2.5-7b \cite{MO-Qwen:qwen2}. 
Table \ref{tab:dif-model} shows the corresponding results of spoofing attacks. 
In Table \ref{tab:dif-model}, OPT-1.3b and OPT-2.7b perform better when $LM_{vic}$=OPT-2.7b than other $LM_{att}$. 
When $LM_{vic}$=Llama3-8b, Llama3-8b exhibits the same advantages. 
This implies that using the same vocabulary can significantly promote the attack effectiveness of SWA. 
We observe that the TPR@1\% of AS is predominantly above 0.9 under identical vocabulary conditions, demonstrating highly effective extraction of the victim watermark. 
Meanwhile, AS consistently achieves better detection metrics than WS, demonstrating its superiority to cross-model stolen.
\label{sec:app-exp-LM_att}

\begin{table}[tbp]
\centering
\scalebox{0.7}{
\begin{tabular}{@{}ccccc@{}}
\toprule
                  & $|ctx|$=1 & $|ctx|$=2 & $|ctx|$=3 & $|ctx|$=4 \\ \midrule
KGW               & \phantom{0}8.65      & \phantom{0}8.68      & \phantom{0}8.63      & \phantom{0}8.63      \\
SynthID           & \phantom{0}9.11      & \phantom{0}9.06      & \phantom{0}9.09      & \phantom{0}9.10       \\
Unbiased          & 10.04     & 10.11     & 12.92     & 12.19     \\
WS                & 10.65     & 11.07     & 10.85     & 11.91     \\
\textbf{AS(Ours)} & 11.69     & 11.15     & 16.92     & 16.34     \\ \bottomrule
\end{tabular}
}
\caption{Execution time (s) of watermarks and attack algorithms under varying $|ctx|$.}
\label{tab:app-effeciency}
\end{table}

\subsection{\texorpdfstring{$|ctx|$}{|ctx|} Analysis}
\label{sec:app-exp-ctx}

$|ctx|$ directly determines the number of different $ctx$ that can be formed, and thus directly affects the complexity of watermarks and SWA. 
However, $|ctx|$ is unknown to the attacker. 
We evaluate spoofing attacks with the victim watermark KGW's $|ctx|$ varying from 1 to 4.
The corresponding experimental results are illustrated in Figure \ref{fig:exp-context-length}. 

The experimental results indicate that AS consistently outperforms WS in terms of stealing effectiveness across various $|ctx|$.
Furthermore, we believe that existing watermarks should enhance $|ctx|$ under permissible conditions. 
Figure \ref{fig:exp-context-length} demonstrates that spoofing attacks are more effective against watermarks with low $|ctx|$ or when the watermark's $|ctx|$ matches the attack configuration ($|ctx|$=3). 
Therefore, the attacker can increase the $|ctx|$ of SWA to ensure the attack is more general, and the victim can also increase the $|ctx|$ of the watermark to enhance defense capability for spoofing attack. 
However, from a practical perspective, the $|ctx|$ of both cannot be infinitely increased. 

\begin{table*}[tbp]
\centering
\scalebox{0.7}{
\begin{tabular}{ccccccccccccccc}
\toprule
     &         &       & \multicolumn{4}{c}{KGW}                              & \multicolumn{4}{c}{SynthID}                          & \multicolumn{4}{c}{Unbiased}                         \\ \cmidrule(lr){4-7} \cmidrule(lr){8-11} \cmidrule(lr){12-15}  
     &         &       & \multicolumn{2}{c}{Dolly} & \multicolumn{2}{c}{Harm} & \multicolumn{2}{c}{Dolly} & \multicolumn{2}{c}{Harm} & \multicolumn{2}{c}{Dolly} & \multicolumn{2}{c}{Harm} \\ \cmidrule(lr){4-5} \cmidrule(lr){6-7} \cmidrule(lr){8-9} \cmidrule(lr){10-11} \cmidrule(lr){12-13} \cmidrule(lr){14-15}
Full & Partial & Empty & WCS          & AUC        & WCS         & AUC        & WCS          & AUC        & WCS         & AUC        & WCS          & AUC        & WCS         & AUC        \\ \midrule
\ding{51}    & \ding{55}       & \ding{55}     & 1.321        & 0.74       & 1.378       & 0.75       & 0.503        & 0.61       & 0.503       & 0.61       & \phantom{-}0.419        & 0.70       & 0.412       & 0.70       \\
\ding{55}    & \ding{55}       & \ding{51}     & 0.373        & 0.54       & 0.385       & 0.56       & 0.500        & 0.49       & 0.499       & 0.45       & -0.022                  & 0.49       & 0.064       & 0.54       \\
\ding{51}    & \ding{51}       & \ding{51}     & 0.994        & 0.70       & 1.056       & 0.70       & 0.502        & 0.58       & 0.502       & 0.56       & \phantom{-}0.237        & 0.63       & 0.291       & 0.65       \\
\ding{55}    & \ding{51}       & \ding{55}     & 0.915        & 0.68       & 1.062       & 0.71       & 0.501        & 0.55       & 0.501       & 0.55       & \phantom{-}0.081        & 0.55       & 0.164       & 0.59       \\
\ding{51}    & \ding{55}       & \ding{51}     & 1.048        & 0.70       & 1.147       & 0.72       & 0.501        & 0.55       & 0.503       & 0.60       & \phantom{-}0.279        & 0.65       & 0.343       & 0.67       \\
\ding{51}    & \ding{51}       & \ding{55}     & 1.160        & 0.72       & 1.281       & 0.74       & 0.502        & 0.58       & 0.503       & 0.60       & \phantom{-}0.288        & 0.65       & 0.314       & 0.65       \\
\ding{55}    & \ding{51}       & \ding{51}     & 0.820        & 0.66       & 0.796       & 0.65       & 0.501        & 0.53       & 0.500       & 0.50       & \phantom{-}0.059        & 0.54       & 0.137       & 0.58       \\ \bottomrule
\end{tabular}
}
\caption{Spoofing attack result on different watermarks with WS as stealing watermark algorithm. The left three columns represent different settings of WS. Higher metric values indicate better attack performance. }
\label{tab:app-steal-ch}
\end{table*}

For SWAs, an increase in $|ctx|$ leads to an increase in the processing complexity of the respective seal.
To evaluate this effect, we measure the inference latency with $|ctx|$ ranging from 1 to 4.
Latency experiment uses Qwen2.5-7b as the attack-assistant model and is conducted on two Tesla-V100 GPUs without concurrent processes. 
Using the c4 dataset described in the main text, we generate 500 samples with 200 new tokens each, and report the average generation time in Table \ref{tab:app-effeciency}. 

As shown in Table \ref{tab:app-effeciency}, for both AS and WS, latency increases due to the growth of $|ctx|$. 
In particular, compared to WS, AS has higher latency. 
Nevertheless, this computational overhead remains acceptable in practice, as AS's execution time is at most twice that of fastest algorithm under identical experimental conditions. 
Regarding the performance degradation of AS when $|ctx|$ increases, we note this is not a critical concern because watermarks do not typically increase $|ctx|$ to extreme values for robustness against spoofing attacks.

\begin{figure}[tbp]
  \centering 
  \scalebox{0.75}{
    \includegraphics[width=\linewidth]{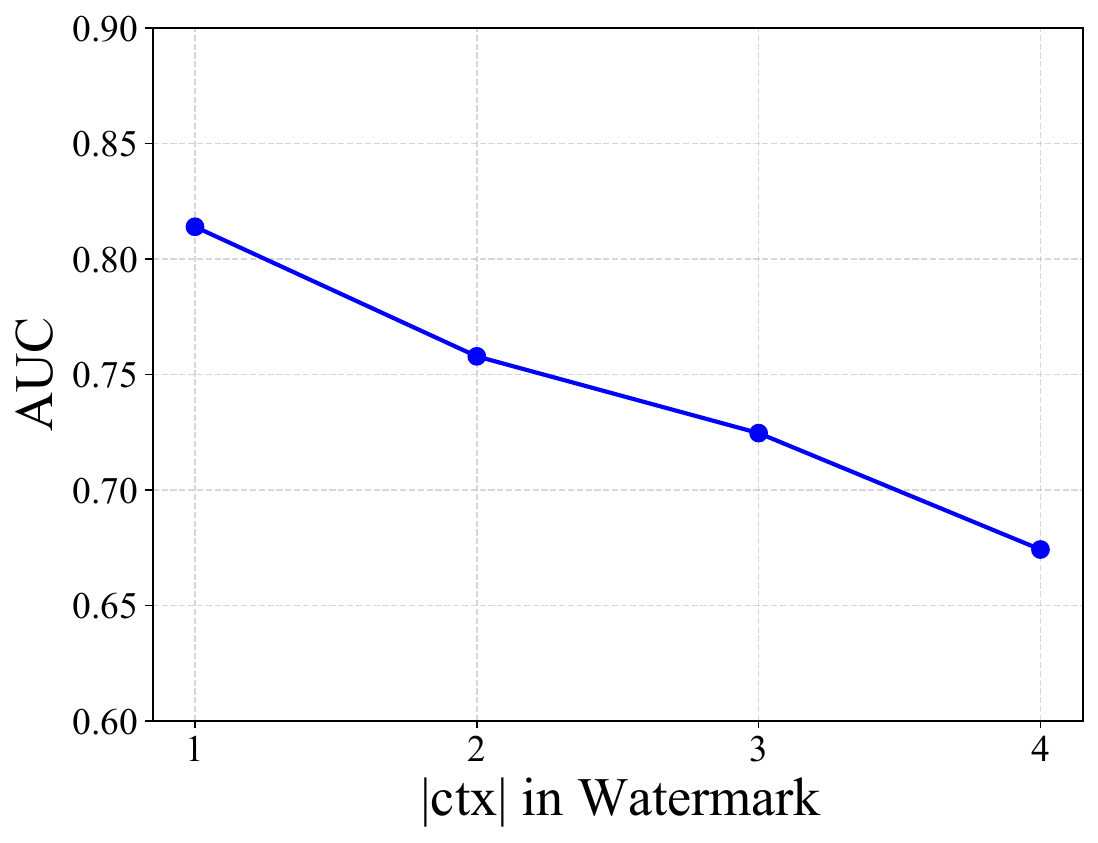}  
  }
  \caption{Results of scrubbing attacks on KGW with different $|ctx|$. The attack method is Dipper, $LM_{vic}$=OPT-2.7b, and dataset is c4.}  
  \label{fig:app-ctx-scrubbing}  
\end{figure}
To further analyze this trade-off, we evaluate how KGW's robustness against scrubbing attacks (using Dipper) changes with varying $|ctx|$, as shown in Figure \ref{fig:app-ctx-scrubbing}.
When KGW faces the same scrubbing attack, the increase in $|ctx|$ causes its robustness to decrease significantly. 
Therefore, when actually deploying the watermark, $|ctx|$ is not set too large. 
This practical constraint naturally limits the $|ctx|$ values that AS needs to handle during attacks.

\subsection{Seal Analsis of WS}

The final seal of WS is obtained by weighted combination of three distinct seals, as formulated in Eq. (\ref{eq:app-steal-weight-score}).
The first seal makes full use of the entire information of $ctx$, and we call it "Full". 
The second seal of WS is named "Partial Context", and we follow its appellation and simplify it to "Partial". 
The third seal of WS ignores all the information of $ctx$, transforms all possible $ctx$ into an empty set, and we call this seal "Empty".
WS sets the weights of (2,1,0.5) for the three seals (Full, Partial, Empty) respectively. 
We follow this weight and conduct experiments on the spoofing attack effect of WS with or without using the three seals. 
In this experiment, we set $LM_{vic}$=Llama3-8b and $LM_{att}$=Qwen2.5-7b.
The experimental results are shown in Table \ref{tab:app-steal-ch}. 

When WS solely employs the "Full" seal, the effect of WS is optimal across three watermarks. 
Whether it is the "Partial" seal (Line 6) or the "Empty" seal (Line 5), both reduce the effect of the pure "Full" seal. 
Therefore, we choose the pure "Full" seal in this paper.

Then it is worth noting that in the work of WS, its authors proved the effectiveness of the "Partial" seal for stealing watermark. 
However, our experiments show that the "Partial" seal is a specialized treatment for the SelfHash-scheme in KGW. 
For the LeftHash-scheme in KGW, as well as SynthID and Unbiased, the "Partial" seal has side effects instead. 
In contrast, the "Full" seal has excellent comprehensive performance to deal with different watermarks. 

Table \ref{tab:app-steal-ch} demonstrates that both the pure "Partial" seal and the "Empty" seal have a certain spoofing attack effect. 
We consider a situation where when executing spoofing attacks to generate tokens, the corresponding $ctx$ may not have been learned by the "Full" seal from $D_w$, resulting in the "Full" seal being unable to be used at this time. 
However, for the "Partial" seal or the "Empty" seal, it can be used when this $ctx$ appears. 
Specifically, for a $ctx$ "two of the", the "Full" seal has only learned the case of "one of the" and cannot recognize "two of the". 
The "Partial" seal can learn the situation "the", and the "Empty" seal can directly give opinions based on the word frequency of the entire $D_w$. 
Therefore, we design Adaptive Selection to dynamically select the seal according to $ctx$, achieving better spoofing effects.

\section{Other Stealing Watermark Algorithms}
\label{sec:app-other-swa}
Besides WS, other scholars also propose different stealing watermark algorithms. 
We briefly introduce them in this section and explain why these algorithms were not compared in the main text. 

\begin{table}[tbp]
\centering
\scalebox{0.7}{
\begin{tabular}{lccccccc}
\hline
                     &            & \multicolumn{2}{c}{KGW} & \multicolumn{2}{c}{SynthID} & \multicolumn{2}{c}{Unbiased} \\ \cline{3-8} 
SWA                  & $LM_{vic}$ & Dolly      & Harm       & Dolly        & Harm         & Dolly         & Harm         \\ \hline
\multirow{2}{*}{CWS} & OPT-2.7B   & 0.450      & 0.514      & 0.537        & 0.513        & 0.501         & 0.520        \\
                     & Llama3-8b  & 0.403      & 0.452      & 0.493        & 0.526        & 0.504         & 0.496        \\ \hline
\multirow{2}{*}{WRA} & OPT-2.7b   & 0.580      & 0.574      & 0.509        & 0.532        & 0.500         & 0.525        \\
                     & Llama3-8b  & 0.544      & 0.572      & 0.531        & 0.526        & 0.522         & 0.484        \\ \hline
\end{tabular}
}
\caption{Spoofing attack results of CWS and WRA. The metric is AUC. }
\label{tab:cws-wra}
\end{table}

\subsection{Common Words Stealing}

\citeauthor{AT-SP-181cww-DBLP:journals/corr/abs-2303-11156} counts the word frequency of 181 common English words, and then judges the possibility score of the $T$ following it being in $\mathcal{G}$ or $\mathcal{R}$.
$\mathcal{G}$ and $\mathcal{R}$ are the green and red lists in the token list partitioning of the KGW watermark. 
In its open-source code, \citeauthor{AT-SP-181cww-DBLP:journals/corr/abs-2303-11156} provides a command-line interface program that guides users to manually select words to evade or steal the victim's watermark. 

We modify its code to enable automatic text generation. 
The score of words by 181 common words is treated as the output of Eq. (\ref{eq:score-funtion}).
When applying impression, we continue to use Eq. (\ref{eq:steal-modify}). 
We define the algorithm Common Words Stealing (CWS). 
Subsequently, we evaluate the effectiveness of spoofing attack using CWS. 
The victim watermark is KGW, and the results are shown in Table \ref{tab:cws-wra}. 

CWS hardly demonstrates effective attacks due to CWS's design covering only the case where $|ctx|=1$. 
Table \ref{tab:cws-wra} shows that the best AUC of CWS is only 0.537, where the worst AUC of WS is 0.59. 
The limited context modeling capability of CWS is incompatible with our experimental watermark configurations that utilize longer context windows. 
Therefore, we do not consider CWS in the main text.

\subsection{Watermark Radioactivity Attack}
Watermark Radioactivity Attack (WRA) \cite{AT-SP-Training-DBLP:conf/acl/PanL0LH0KY25} is a fine-tuning-based implementation of SWA. 
Fine-tuning is a common approach for models to learn data patterns, and can also be used to learn (or steal) watermark information from watermarked text.

However, compared to approaches like WS and AS that are based on token statistical reasoning, fine-tuning has higher requirements for both attack resources and data. 
During testing, two Tesla-V100 GPUs only supported fine-tuning of OPT-1.3b, and the spoofing attack results are presented in Table \ref{tab:cws-wra}.
WRA achieves limited success in spoofing attacks with AUC values consistently below 0.6.
Compared to WS and AS in Table \ref{tab:dif-wm}, WRA significantly underperforming AS and WS.

With substantially more training data, WRA might potentially generate more fluent watermarked text than statistical approaches like WS and AS.
However, in real-world attack scenarios, adversaries typically face the Limited Queries constraint, where only a small number of watermarked samples can be obtained from the victim model. 
This means that AS and WS have higher attack performance than WRA in practical attack scenarios. 
Both in terms of computational resource requirements and attack effectiveness under Limited Queries constraints, we relegate detailed analysis of WRA this section, focusing our main evaluation on more practical attack methods.

\subsection{Mixed Integer Programming}
\begin{table}[tbp]
\centering
\scalebox{0.7}{
\begin{tabular}{@{}lcccc@{}}
\toprule
         & \multicolumn{2}{c}{$LM_{vic}$=OPT-2.7b} & \multicolumn{2}{c}{$LM_{vic}$=Llama3-8b} \\
         & Dolly              & Harm               & Dolly               & Harm               \\ \midrule
KGW      & 0.977              & 0.902              & 0.970               & 0.899              \\
SynthID  & 0.982              & 0.897              & 0.938               & 0.855              \\
Unbiased & 0.976              & 0.905              & 0.939               & 0.860              \\ \bottomrule
\end{tabular}
}
\caption{Scrubbing attack results of MIP. $LM_{att}$=Dipper and metric is AUC. Lower AUC values indicate better watermark removal performance.}
\label{tab:dp-mip}
\end{table}
Mixed Integer Programming (MIP) formalizes the spoofing attack as a mixed integer programming problem with constraints, and then predict more accurate $\mathcal{G}$ tokens \cite{AT-SP-greenlistPredict-DBLP:conf/acsac/ZhangZZZ0HGP24}. 

MIP focuses more on the scenario where watermark sets multiple keys, with Unigram \cite{WM2-L-Unigram-DBLP:conf/iclr/ZhaoA0W24} as the main attack watermark. 
Meanwhile, Unigram is essentially the extreme case of KGW when $|ctx| = 0$.

MIP has only open-sourced the code for conducting scrubbing attacks. 
We evaluates its scrubbing performance, with results presented in Table \ref{tab:dp-mip}. 
When compared with WS and AS in Table \ref{tab:scrubbing-table}, MIP exhibits significantly lower attack effectiveness.

MIP's disadvantage is identical to CWS, as its exclusive focus on low $|ctx|$ makes it difficult to handle complex watermark settings.
Moreover, MIP's open-source implementation lacks support for spoofing attack.
For these reasons, MIP is not included in our main analysis.

\subsection{De-Mark}
De-Mark\cite{AT-SP-Demark:DBLP:conf/icml/ChenWGH25} is a special SWA, and its core highlight lies in using adversarial prompts to determine the $\hat{im}_T$ of the specific $T$ after $ctx$.
However, both AS and other SWAs use conventional watermark text for watermark information extraction. 
This makes it impossible to achieve a fair evaluation between De-Mark and other SWAs. 
Therefore, this paper does not consider comparing De-Mark. 

In terms of actual attack effectiveness, De-Mark and other SWAs have different focuses. 
The core limitation of all SWA algorithms is the number of queries.
De-Mark can generate several highly detectable spoofing attack texts under this limitation, but continuing to generate more is limited by the already recorded impression $\hat{im}$ of $ctx$. 
Other SWAs can extract more impressions of $ctx$ than De-Mark in the same queries, but low-frequency $ctx$ or unrecorded $ctx$ will introduce noise during generation.
From a certain perspective, our AS suppresses the impact of this noise, thereby enhancing the aggressiveness. 

\section{Ethical Considerations and Potential Risks}
AS, while intended for defensive research purposes, carries potential risks that warrant discussion. 

\noindent\textbf{Misuse Potential}: The techniques described could be misused to circumvent watermark detection systems designed to identify AI-generated content, potentially enabling malicious actors to distribute deceptive content at scale.

\noindent\textbf{Stakeholder Impact}: Content platforms, educators, and users relying on watermark detection for content authenticity verification could be negatively impacted if our methods are deployed without appropriate safeguards.

\noindent\textbf{Mitigation Strategy}: To minimize misuse risk, we: (1) actively engage with watermark developers to strengthen their systems; and (2) recommend that future watermark designs incorporate robustness against AS. 

\noindent\textbf{Dual Use Consideration}: While AS could weaken existing watermark systems, its primary purpose is to proactively identify vulnerabilities before malicious actors do. We believe this defensive research ultimately strengthens the ecosystem by enabling more robust watermark designs that can withstand sophisticated attacks.

\noindent\textbf{Data Safety}: 
We use three main datasets: (1) C4 (Colossal Clean Crawled Corpus), which contains English web text across diverse domains, filtered for cleanliness and deduplication \cite{DS-C4:journals/jmlr/RaffelSRLNMZLL20}; (2) Dolly, a dataset of human-generated instruction-response pairs covering multiple domains \cite{DS-Dolly:DatabricksBlog2023DollyV2}; and (3) Harm, a dataset of harmful prompts derived from AdvBench \cite{DS-advbench:DBLP:journals/corr/abs-2307-15043} and HarmfulQ \cite{DS-harmfulq:DBLP:conf/acl/Shaikh0HBY23}. 
Our experiments focus exclusively on English data, limiting applicability to other languages. The demographic characteristics of content creators in these datasets reflect the biases of their source materials (primarily web content), which may impact generalizability across different populations.
The C4 dataset has been preprocessed to remove personally identifiable information. The Harm dataset, constructed from HarmfulQ and AdvBench, intentionally contains harmful content for evaluation purposes. 
We use this dataset solely for evaluating watermark removal against harmful content generation and do not further process the harmful content as it is essential for the evaluation scenario. 

\section{Defend Adaptive Stealing}
Besides encouraging future scholars to conduct research based on AS, we have summarized several feasible defense strategies according to existing conclusions. 

First, increase the $|ctx|$ of the watermark. 
According to Appendix \ref{sec:app-exp-ctx}, an increase in $|ctx|$ substantially raises the difficulty for the AS to compromise the watermark. 

The second approach is to employ sentence-level watermarks. 
The AS focuses on watermarking methods that embed information during token generation, with tokens serving as the embedding units. 
Sentence-level watermarking expands the embedding unit to encompass a single sentence. 
This discrepancy causes a mismatch between the AS's stealing target and the actual implementation, thereby rendering the attack ineffective.

Adopting a more complex hashing scheme may represent a mandatory approach for watermark deployers. 
When the hashing schemes employed by all watermarking methods utilize every token within the context for mapping, the theoretically strongest forged seal corresponds to an $[1,1,...,1]$ seal. 
This configuration consequently confines the attack performance of the AS strictly to a level bounded by this seal.

Random key selection is also a feasible defense method, but this method is often accompanied by a decrease in the performance and efficiency of detection.
\end{document}